\documentclass[%
 reprint,
 amsmath,amssymb,
 aps, prab,
]{revtex4-2}

\usepackage{graphicx}
\usepackage{dcolumn}
\usepackage{bm}

\newcommand{\co}{_{\textrm{c.o.}}}
\newcommand{\degree}{^\circ}
\newcommand{\mat}[1]{\mathsf{#1}}  
\renewcommand\vec{\bm}
\newcommand{\hatvec}[1]{\vec{\hat{{#1}}}}
\DeclareMathOperator*{\argmax}{argmax}

\newcommand{\Lagr}{\mathcal{L}}

\begin{document}

\preprint{APS/123-QED}

\title{BAGELS for simultaneous polarization, orbit,\\ and optics control in electron storage rings}%
\thanks{Work supported by Brookhaven Science Associates, LLC under Contract No. DE-SC0012704 with the U.S. Department of Energy.}%

\author{M. G. Signorelli}
\email{mgs255@cornell.edu}
\author{G. H. Hoffstaetter}%
  \altaffiliation[Also at ]{Brookhaven National Laboratory}
\affiliation{%
 Department of Physics, Cornell University, Ithaca, NY, USA
}%

\date{\today}

\begin{abstract}
We present a new method for minimizing the effects of radiative depolarization in electron storage rings by use of a minimal number of special vertical orbit bumps. The bumps can be used to minimize the effects of radiative depolarization while simultaneously maintaining other common benefits of vertical orbits, e.g. transverse coupling and vertical dispersion control. Because simultaneously optimizing the large number of vertical correctors in a ring is operationally infeasible, we use dimensionality reduction to define a minimal number of most effective groups of vertical correctors that can be optimized during operation, motivating the name ``Best Adjustment Groups for ELectron Spin'' (BAGELS). The method is streamlined by using suitable ``basis bumps'' instead of all individual vertical correctors. We define three types of basis bumps for different purposes: (1) generates no delocalized transverse coupling nor delocalized vertical dispersion, (2) generates no delocalized vertical dispersion, and (3) generates no delocalized transverse coupling. BAGELS has been essential in the design of the Electron Storage Ring (ESR) of the Electron-Ion Collider (EIC), and will be beneficial for any polarized electron ring, including FCC-ee. HERA and LEP would have likely benefitted as well. We use BAGELS to significantly increase polarization in the 18 GeV EIC-ESR, beyond achievable with conventional methods; in the 1-IP lattice, we nearly double the asymptotic polarization, and in the 2-IP lattice we more than triple the asymptotic polarization. We also use BAGELS to construct knobs that can be used for global coupling correction, and knobs that generate vertical emittance for beam size matching, all while having minimal impacts on the polarization and orbit/optics.
\end{abstract}

\maketitle


\section{Introduction}
Spin-polarized particle colliders provide one of the best probes of the internal structures of fundamental particles and their interactions, and remain one of the most important tools in high energy physics. The Hadron-Elektron-RingAnlage (HERA) collider at DESY, which operated between 1992-2007, collided protons with spin-polarized electrons or positrons, providing valuable high energy physics results, including excellent experimental agreement of the electro-weak interaction spin dependence with standard model predictions \cite{Abramowicz2010}. Now, the Electron-Ion Collider (EIC) soon to be built at Brookhaven National Laboratory will push the frontier even further, being the first circular collider of longitudinally-spin-polarized electrons with longitudinally-spin-polarized light-ions for a wide range of selected center-of-mass energies \cite{osti_1765663}. Data from such collisions in the EIC will answer important questions about the origin of proton spin, among others.

\begin{figure}[b]
   \centering
   \includegraphics*[width=1\columnwidth]{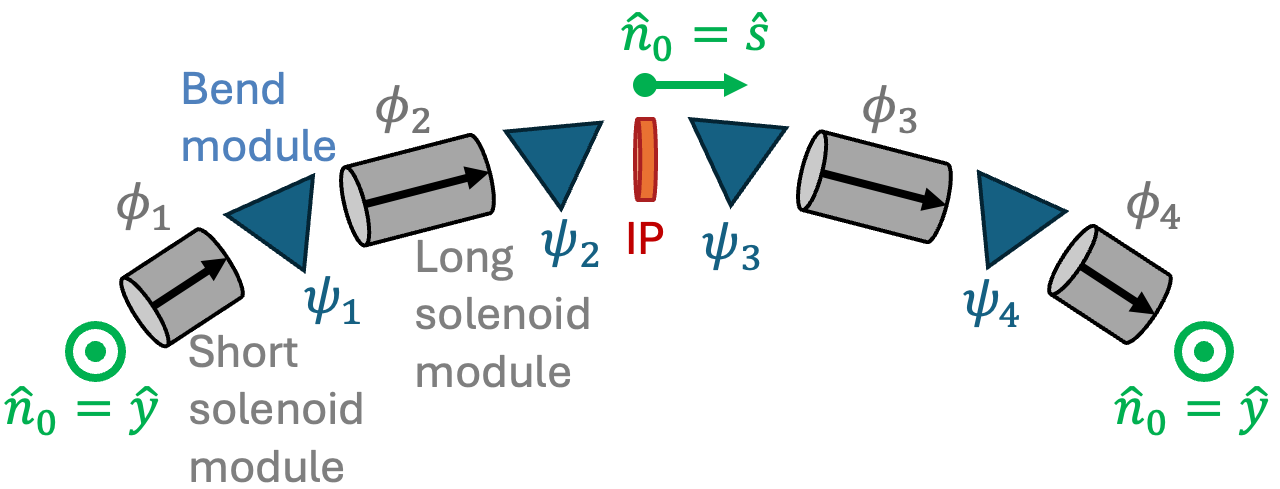}
   \caption{Top-down view of the spin rotator of the ESR, with the spin precession angles in each module labelled.}
   \label{fig:esr-rotator}
\end{figure}

Maintaining high spin polarization throughout storage for sufficiently long amounts of times is essential to achieving optimal physics data. The Electron Storage Ring (ESR) of the EIC aims to provide longitudinally-polarized electron bunches at each of roughly 5, 10 and 18 GeV in both a 1- and 2-colliding interaction point (IP) configuration using the spin rotator shown in Fig.~\ref{fig:esr-rotator}. For each energy (which defines the bend spin precession $\psi_j$), the solenoid strengths $\phi_i$ are set for longitudinal spin at the IP. In electron rings, the polarization evolution is dominated by the stochastic emission of synchrotron radiation. In a ring with spin rotators, such the ESR, radiative depolarization will rapidly reduce the polarization of a bunch unless remedied by \textit{strong synchro-beta spin matching} - choosing the magnet strengths and spin rotator configuration so that the 1-turn spin dependence on the orbit motion is removed at as many azimuths as possible \cite{deshandbook,chao}. The ESR satisfies a horizontal strong synchro-beta spin match everywhere outside of the spin rotator, however, a longitudinal strong synchro-beta spin match is not possible due to technical constraints; 11 Tesla solenoids would be required, and geometry changes are restricted by the current RHIC tunnel.

While the polarization requirements are still satisfied for the 5 and 10 GeV cases, at 18 GeV the lack of a longitudinal spin match is catastrophic, resulting in marginally sufficient polarization in the ideal 1-IP lattice and insufficient polarization in the ideal 2-IP lattice \cite{Signorelli:2023fqc}. Such ideal lattices also do not yet include any vertical emittance creation schemes, necessary to match the electron beam size with the ion beam size for maximum luminosity. Without a careful approach to increasing the vertical emittance, the polarization is all-but-guaranteed to degrade even more. As a result, a new, feasible, minimally-invasive method for reducing the effects of radiative depolarization in the ideal lattice, the real ring including errors, and the ring with sufficient vertical emittance, was necessary.

Here we present such a solution, which we call ``Best Adjustment Groups for ELectron Spin" (BAGELS), that minimizes the radiative depolarization by using a minimal number of special vertical closed orbit bumps. BAGELS provides a rigorous framework for calculating vertical closed orbit bumps which maximally impact the so-called \textit{spin-orbit coupling function} around the ring, while having a minimal impact on the orbit and by construction generating no delocalized vertical dispersion nor transverse coupling (caused by the vertical orbit offsets in the sextupoles). Knobs controlling these special bumps can be used to create spin-orbit coupling which cancels out that leftover by the rotator in the ideal lattice, and that created by the random errors in the real ring. This approach, which directly attacks the cause of radiative polarization while creating minimal orbit excursions, effectively makes harmonic closed orbit spin matching obsolete for correcting polarization in operation. And, BAGELS can also be applied in applications where a minimal impact on polarization is desired. For example, with BAGELS we can calculate a minimal number of knobs to optimally correct the global coupling caused by random errors, while ensuring such knobs have minimal impacts to the spin-orbit coupling function and closed orbit. We also can use BAGELS to safely generate vertical emittance by constructing a special vertical orbit bump that either maximally creates only delocalized coupling (to couple the horizontal emittance into the vertical) or maximally creates only delocalized vertical dispersion (to excite a vertical amplitude when a photon is emitted), all while having minimal impacts to both the spin-orbit coupling function and the closed orbit itself. 

This paper is organized as follows: in Sec.~\ref{sec:theory} we outline in-depth the theory of electron polarization in storage rings, derive the equations of motion for first-order spin-orbit motion, and describe traditional methods used (e.g. in HERA) to minimize the effects of radiative depolarization. In Sec.~\ref{sec:methods} we describe the computational methodology employed for calculating the various quantities and details of all Monte Carlo spin tracking simulation results presented in this paper. In Sec.~\ref{sec:bagels} we rigorously formulate BAGELS, describing the physical mechanisms of how it works and how to apply it in different applications. Finally in Sec.~\ref{sec:results} we show how BAGELS applied in the 18 GeV EIC-ESR to spin match the ideal 1-IP and 2-IP ESR, to correct the spin match degradation (and delocalized coupling) caused by random errors for 10 different 1-IP ESR seeds, and to generate sufficient using either delocalized coupling or delocalized vertical dispersion with minimal impacts to the polarization.




%

\section{Theory}\label{sec:theory}
\subsection{Electron Bunch Polarization Time Evolution}
To analyze the spin dynamics and polarization evolution in a periodic accelerator, the calculation of a spin invariant is required. The so-called \textit{invariant spin field (ISF)} $\hatvec{n}$ is a special 1-turn periodic spin field that solves the spin equation of motion along orbital trajectories. Explicitly,

\begin{equation}\label{eq:ISF}
        \hatvec{n}(\vec{M}(\vec{z}_i)) = \mat{R}(\vec{z}_i)\hatvec{n}(\vec{z}_i) \ ,
\end{equation}

where we choose $\vec{z}_i=(x,p_x,y,p_y,z,\delta)_i^T$ to be the initial canonical phase space coordinates in a coordinate system with its origin on the closed orbit (i.e. $\vec{z}\co=\vec{0}$), $\delta$ is the relative momentum deviation from the closed orbit momentum $p_0$, $\vec{M}$ is the 1-turn map, and $\mat{R}$ is the spin transport rotation matrix for a particle through one turn. Note we have omitted the azimuthal position around the ring for brevity. The ISF, when it exists, is unique up to a flip in sign ($\hatvec{n}$ and $-\hatvec{n}$ both satisfy Eq.~\eqref{eq:ISF}), and the ISF evaluated on the closed orbit $\hatvec{n}(\vec{z}\co)$ is referred to as $\hatvec{n}_0$ which, unlike for general orbits, can be obtained from the eigenvector with unit eigenvalue of the 1-turn spin rotation matrix on the closed orbit. The projection of any particle's spin along the ISF is called the spin action, $J_s=\vec{S} \cdot \hatvec{n}$. The spin action is an adiabatic invariant, in that if some dynamical parameter is varied slowly with respect to the orbital angle evolution, $J_s$ remains constant \cite{isf}. The ultimate goal in designing spin polarized rings is to minimize the mechanisms and rates at which spin actions in a beam may be reduced.

For electrons in storage rings, bunch polarization evolution is largely dominated by the stochastic emission of synchrotron radiation, which causes three effects: (1) the \textit{Sokolov-Ternov (ST) effect} which is an asymmetry in the spin flip rate during photon emission leading to a buildup of polarization antiparallel to the bending field for electrons at a rate $\tau_{st}^{-1}$, (2) \textit{radiative depolarization} where the instantaneous jump in phase space coordinates after photon emission changes a particle's spin action and rapidly reduces the polarization of a bunch at a rate $\tau_{dep}^{-1}$, and (3) \textit{kinetic polarization} where the ISF at the resulting phase space coordinate after photon emission may actually be more aligned with a particle's spin than before \cite{st,bk,bks,dk}. The effect of kinetic polarization is only significant when $\hatvec{n}_0$ is not vertical in significant portions of the ring. All three of these effects balance each other out over time for leading to an asymptotic polarization ultra-relativistic electrons given by the Derbenev-Kondratenko (DK) formula,

\begin{equation}\label{eq:Pdk}
    P_{dk} = \frac{8}{5\sqrt{3}}\frac{\oint ds \left\langle |\dot{\hatvec{v}}|^3 \ \hatvec{b}\cdot\left(\hatvec{n} - \vec{d}\right)\right\rangle}{\oint ds \left\langle |\dot{\hatvec{v}}|^3\left( 1 - \frac{2}{9}(\hatvec{n}\cdot\hatvec{v})^2 + \frac{11}{18}\left\lvert\vec{d}\right\rvert^2\right) \right\rangle} \ ,
\end{equation}

where $\hatvec{v}$ is a unit vector in the direction of the particle's velocity, $\dot{\hatvec{v}}$ is the curvature of the particle trajectory, $\hatvec{b}=\hatvec{v}\times\dot{\hatvec{v}}/|\dot{\hatvec{v}}|$, $\langle ... \rangle$ denotes an average over the particle ensemble, $s$ is the arc position along the closed orbit, and $\vec{d}=\gamma_0\frac{\partial\hatvec{n}}{\partial\gamma}$ is the so-called \textit{spin-orbit coupling function}. The spin-orbit coupling function captures the dependence of the ISF with energy, and thus defines the degree of radiative depolarization present in a ring. The DK formula is only accurate to first-order in energy deviations from the closed orbit energy $\gamma_0$, and also explicitly approximates the electron/positron anomalous magnetic moment $a \approx 0$. However as shown rigorously by Mane \cite{mane-PhysRevA.36.105}, assuming $a\approx 0$ is generally not appropriate in the implicit calculations of $\hatvec{n}$ and $\vec{d}$. In Eq.~\eqref{eq:Pdk}, the $\hatvec{b}\cdot\vec{d}$ term corresponds to the kinetic polarizing mechanism, $\frac{11}{18}\lvert \vec{d}\rvert^2$ term to the radiative depolarization, and the remaining terms to the Sokolov-Ternov effect. The time evolution of a bunch's polarization is
\begin{equation}\label{eq:Pt}
    P(t) = P_{dk}(1-e^{-t/\tau_{dk}})+P_0e^{-t/\tau_{dk}}
     \ , 
     \end{equation}
where $\tau_{dk}^{-1} = \tau_{st}^{-1} + \tau_{dep}^{-1}$, and $P_0$ is the initial bunch polarization. The Sokolov-Ternov rate and radiative depolarization rate are respectively

\begin{align}\label{eq:taust}
    \tau_{st}^{-1}&=\frac{5\sqrt{3}}{8}\frac{r_e\gamma_0^5\hbar}{m_e}\frac{1}{C}\oint ds \left\langle |\dot{\hatvec{v}}|^3\left(1-\frac{2}{9}(\hatvec{n}\cdot\hatvec{v})^2\right)\right\rangle \ , \\\label{eq:taudep}
    \tau_{dep}^{-1}&=\frac{5\sqrt{3}}{8}\frac{r_e\gamma_0^5\hbar}{m_e}\frac{1}{C}\oint ds \left\langle |\dot{\hatvec{v}}|^3\frac{11}{18}\left\lvert\vec{d}\right\rvert^2\right\rangle  \ ,
\end{align}

where $r_e$ is the classical electron radius, $m_e$ is the electron mass, and $C$ is the circumference of the ring.

For electron rings, the primary task is to reduce $\vec{d}$ around the ring as much as possible, thus minimizing the radiative depolarization rate and maximizing the asymptotic polarization. In a perfectly midplane symmetric ring without spin rotators, $\vec{d}$ is already zero everywhere; in such a case the vertical beam size is approximately zero due to radiation damping, and the beam (which lies entirely in the midplane) only sees vertical magnetic fields. The ISF is therefore fully vertical in all occupied areas of phase space. However, once $\hatvec{n}_0$ is rotated out of the vertical (e.g. by spin rotators or random errors), $\vec{d}$ will be excited and cause rapid radiative depolarization if left uncorrected. 

\subsection{First-Order Spin-Orbit Motion}
To understand the dominant physical mechanisms which enable BAGELS to work, it is insightful to consider the theory in a simplified model. In this section, we invoke the paraxial approximation and consider the dynamics to first-order in deviations from the closed orbit. Using a coordinate system which has its origin on the closed orbit ($\vec{z}\co=\vec{0}$), we linearize in deviations of the phase space coordinates $\vec{\Delta}(\vec{z}\co=\vec{0};\vec{z})=\vec{z}-\vec{z}\co=(x,p_x,y,p_y,z,\delta)^T$. We also linearize in deviations of the trace space coordinates $x'$, $y'$, where $'$ denotes differentiation with respect to the arc position along the closed orbit $s$. We define the right-handed orthonormal coordinate system $(\hatvec{x}(s),\hatvec{y}(s),\hatvec{s}(s))$ along the closed orbit (note that $\hatvec{y}$ may not always point vertically with respect to the geometric layout of the accelerator, for example). The local horizontal and vertical curvatures of the coordinate system are specified by $\vec{g}=g_x\hatvec{x}+g_y\hatvec{y}$. In the absence of electric fields, by definition $g_x = \frac{q}{p_0}B_y\rvert_{\vec{0}}$ and $g_y=-\frac{q}{p_0}B_x\rvert_{\vec{0}}$ where $\vec{B}$ is the magnetic field in the closed orbit basis. For example, $\vec{g}$ will be nonzero when the closed orbit through a quadrupole magnet is offset from the magnet centerline. Note that with this definition $\vec{g}$ points in the opposite direction of $\dot{\hatvec{v}}$. 
The Thomas-BMT equation in this coordinate system is

\begin{equation}
\frac{d\vec{S}}{dt} =\vec{\Omega}_t\times\vec{S}  \nonumber \ , 
\end{equation}
\begin{eqnarray}
\vec{\Omega}_t =&& -\frac{q}{p}v \left[(1+G\gamma)\vec{B} - G(\gamma-1)(\hatvec{\beta}\cdot\vec{B})\hatvec{\beta}\right. \nonumber \\
&&\left.+ \left(G\gamma\beta+\frac{\gamma\beta}{1+\gamma}\right)\frac{\vec{E}\times\hatvec{\beta}}{c}\right] -\frac{ds}{dt} \vec{g}\times\hatvec{s} \ ,\label{eq:TMBT1}
\end{eqnarray}

where $\vec{S}$ is a 3-vector of the spin expectation values in the closed orbit basis, $G$ is the anomalous magnetic moment of the particle, and $\vec{B}$ and $\vec{E}$ are the laboratory frame magnetic and electric fields in the closed orbit basis \cite{thomas1927, BMT, thomas1982, georg}. The $\vec{g}\times\hatvec{s}$ term accounts for the rotation of the coordinate system. Expanding around the closed orbit (letting $\vec{r}_t = x\hatvec{x}+y\hatvec{y}$),

\begin{align}    
    v&\approx \frac{ds}{dt}\left(1+\vec{g}\cdot\vec{r}_t\right) \ , \label{eq:v}\\
    \frac{q}{p} &\approx \frac{q}{p_0}(1-\delta) \ , \label{eq:qp}\\
    \gamma&\approx \gamma_0(1+\beta_0^2\delta)\ , \label{eq:gamma} \\
    \beta &\approx \beta_0\left(1+\delta/\gamma_0^2\right) \ , \label{eq:beta}\\    
    \hatvec{\beta} &\approx (x'\hatvec{x}+y'\hatvec{y}+\hatvec{s}) \ ,\label{eq:betahat}\\
    \vec{p} &\approx p_0\left[x'\hat{x}+y'\hat{y}+(1+\delta)\hat{s}\right] \ . \label{eq:p}
\end{align}

Note that $\frac{ds}{dt}$ implicitly contains a $\delta$ dependence. Multiplying both sides of Eq.~\eqref{eq:TMBT1} by $\frac{dt}{ds}$ and substituting Eqs.~\eqref{eq:v} through \eqref{eq:betahat}, the Thomas-BMT equation can be rewritten as

\begin{equation}
    \vec{S}'= \vec{\Omega}\times \vec{S}\label{eq:TBMT}
\end{equation} where $\vec{\Omega}=\vec{\Omega}_B+\vec{\Omega}_E - \vec{g}\times\hatvec{s}$ and

\begin{widetext}
\begin{subequations} \label{eq:Omega_B}
  \begin{eqnarray}
    \Omega_{B,x} \approx &&-\frac{q}{p_0}\left\lbrace\left[(1+G\gamma_0)(1 + \vec{g}\cdot\vec{r}_t) - \left(1+\frac{G}{\gamma_0}\right)\delta\right]B_x + (G-G\gamma_0)x' B_s\right\rbrace \ ,\\
    \Omega_{B,y} \approx &&-\frac{q}{p_0}\left\lbrace\left[(1+G\gamma_0)(1 + \vec{g}\cdot\vec{r}_t) -\left(1+\frac{G}{\gamma_0}\right)\delta\right]B_y +(G-G\gamma_0)y'B_s\right\rbrace\ ,\\
    \Omega_{B,s} \approx &&-\frac{q}{p_0}\Biggl\lbrace (1+G)(1+ \vec{g}\cdot\vec{r}_t - \delta)B_s+(G - G\gamma_0)(x'B_x + y'B_y)
    \Biggr\rbrace \ ,
    \end{eqnarray}\end{subequations}\begin{subequations}
    \begin{eqnarray}
   \Omega_{E,x} \approx&&-\frac{q}{p_0}\left\lbrace\left[G\beta_0\gamma_0+ \frac{\beta_0\gamma_0}{1+\gamma_0}\right]\left[(1+\vec{g}\cdot\vec{r}_t)\frac{E_y}{c} - y'\frac{E_s}{c}\right] + \frac{\beta_0\gamma_0}{1+\gamma_0}\left[\frac{1}{\gamma_0}-1\right]\delta\frac{E_y}{c}\right\rbrace \ ,\\
   \Omega_{E,y} \approx && -\frac{q}{p_0}\left\lbrace\left[G\beta_0\gamma_0+\frac{\beta_0\gamma_0}{1+\gamma_0}\right]\left[x'\frac{E_s}{c} - (1+\vec{g}\cdot\vec{r}_t)\frac{E_x}{c}\right] + \frac{\beta_0\gamma_0}{1+\gamma_0}\left[1-\frac{1}{\gamma_0}\right]\delta\frac{E_x}{c}\right\rbrace \ ,\\
   \tilde \Omega_{E,s} \approx &&-\frac{q}{p_0}\left\lbrace\left[G\beta_0\gamma_0+\frac{\beta_0\gamma_0}{1+\gamma_0} \right]\left[y'\frac{E_x}{c} - x'\frac{E_y}{c}\right]\right\rbrace \ .
\end{eqnarray}  
\end{subequations}
\end{widetext}

We now consider only magnetostatic fields, expanded around the closed orbit to first-order. For convenience we define the quantities

\begin{align}
    K_s &= \frac{q}{p_0}\left.B_s\right\rvert_{\vec{0}} \ ,\\
    K_1&= \frac{q}{p_0}\left.\partial_{y }B_x\right\rvert_{\vec{0}} \ , \\ 
    \tilde{K}_1 &= \frac{q}{2p_0}\left(\left.\partial_xB_x\right\rvert_{\vec{0}}-\left.\partial_yB_y\right\rvert_{\vec{0}}\right) \ .
\end{align}

In the special case where the closed orbit is aligned with the magnet centerlines, these quantities are equivalent to the normalized solenoid, quadrupole, and skew quadrupole strengths respectively. The general magnetic field satisfying the Maxwell equations $\vec{\nabla}\times\vec{B}=\vec{\nabla}\cdot\vec{B}=0$ to first-order is

\begin{subequations}\label{eq:B}
\begin{eqnarray}
    B_x \approx&& -\frac{p_0}{q}\left(\frac{1}{2}K_s'x +  g_y - K_1y + \tilde{K}_1x \right), \\
    B_y \approx&& -\frac{p_0}{q}\left(\frac{1}{2}K_s'y-g_x-K_1x-\tilde{K}_1y\right), \\
    B_s \approx&& -\frac{p_0}{q}\left(- K_s-g_x'y+g_y'x \right)\ . 
\end{eqnarray}
\end{subequations}

Substituting Eq.~\eqref{eq:B} into Eq.~\eqref{eq:Omega_B}, we split $\vec{\Omega}$ into a zeroth order part $\vec{\Omega}_0$ on the closed orbit and a perturbative part $\vec{\omega}(\vec{\Delta})$ first-order in the phase space variables,

\begin{subequations} \label{eq:Omega}
\begin{align}
    \Omega_{0,x} &\approx  G\gamma_0g_y  \ , \\
    \Omega_{0,y} &\approx -G\gamma_0g_x\ , \\
    \Omega_{0,z} &\approx -(1+G)K_s  \ ,
\end{align}    
\end{subequations}

\begin{widetext}
    \begin{subequations}\label{eq:omega_b1}
\begin{align}
    \omega_x &\approx \left(1+G\gamma_0\right)\left(\frac{1}{2}K_s'x+g_y\vec{g}\cdot\vec{r}_t-K_1y+\tilde{K}_1x\right)+\left(G\gamma_0-G\right)K_sx'-\left(1+\frac{G}{\gamma_0}\right)g_y\delta \ , \label{eq:omega_x}\\
    \omega_y &\approx \left(1+G\gamma_0\right)\left(\frac{1}{2}K_s'y-g_x\vec{g}\cdot\vec{r}_t-K_1x-\tilde{K}_1y\right)+\left(G\gamma_0-G\right)K_sy'+\left(1+\frac{G}{\gamma_0}\right)g_x\delta \ ,\label{eq:omega_y}\\
   \omega_s &\approx \left(1+G\right)\left(-K_s\vec{g}\cdot\vec{r_t} + K_s\delta
   -g_x'y+g_y'x\right)+\left(G\gamma_0-G\right)\left(g_xy'-g_yx'\right) \ .
\end{align}
\end{subequations}
\end{widetext}

Equivalent, alternative expressions for $\omega_x$ and $\omega_y$ can be obtained by substituting in the orbital equations of motion. Substituting Eqs.~\eqref{eq:v},~\eqref{eq:betahat},~\eqref{eq:p} and \eqref{eq:B} into $\frac{d\vec{p}}{dt}=qv\hatvec{\beta}\times\vec{B}$, we obtain

\begin{equation}
       x''-K_sy'-g_x\delta \approx \frac{1}{2}K_s'y-g_x\vec{g}\cdot\vec{r}_t-K_1x-\tilde{K}_1y \ , \label{eq:eom1}
\end{equation}

   \begin{equation}
         -y''-K_sx'+g_y\delta \approx \frac{1}{2}K_s'x+g_y\vec{g}\cdot\vec{r}_t-K_1y+\tilde{K}_1x\ . \label{eq:eom2}  
   \end{equation}

Substituting Eq.~\eqref{eq:eom1} into Eq.~\eqref{eq:omega_y} and Eq.~\eqref{eq:eom2} into Eq.~\eqref{eq:omega_x} gives 

\begin{widetext}
\begin{subequations}\label{eq:omega_b2}
    \begin{align}
            \omega_x &\approx -\left(1+G\gamma_0\right)y''-\left(1+G\right)K_sx'+\left(G\gamma_0-\frac{G}{\gamma_0}\right)g_y\delta \ ,\\
    \omega_y &\approx\left(1+G\gamma_0\right)x''-\left(1+G\right)K_sy'-\left(G\gamma_0-\frac{G}{\gamma_0}\right)g_x\delta \ .
    \end{align}
\end{subequations}
\end{widetext}



\subsection{Traditional Methods to Minimize Radiative Depolarization}
To minimize the effects of radiative depolarization, a \textit{strong synchro-beta spin match} should first be applied \cite{deshandbook,chao}: the magnet strengths and spin rotator configuration are chosen so that the 1-turn spin dependence on the orbit motion is removed at as many azimuths as possible. By minimizing the 1-turn spin-orbit coupling where photons are emitted, $\vec{d}$ is also minimized. This is usually done to first-order in the phase space coordinates' deviation from the closed orbit. To do so, we first construct a right-handed orthonormal coordinate system $(\hatvec{n}_0,\hatvec{m}_0,\hatvec{l}_0)$ where $\hatvec{l}_0$ and $\hatvec{m}_0$ solve the Thomas-BMT equation along the closed orbit. Note that this coordinate system is not 1-turn periodic, with $\hatvec{m}_0$ and $\hatvec{l}_0$ having precessed by the closed orbit spin tune $\nu_0$ around $\hatvec{n}_0$ after one turn. We can approximate any particle's spin for small deviations as $\hatvec{n}_0$ as 
\begin{equation} \label{eq:SLIMS}
    \vec{S}\approx\hatvec{n}_0+\zeta_1\hatvec{m}_0+\zeta_2\hatvec{l}_0 \ ,
\end{equation}
to first-order in $\zeta_1$ and $\zeta_2$. Substituting Eq.~\eqref{eq:SLIMS} into Eq.~\eqref{eq:TBMT}, we obtain 

\begin{equation}
    \frac{d\zeta_1}{ds} \approx \vec{\omega}(\vec{\Delta}(\vec{z}))\cdot\hatvec{l}_0 \ , \ \frac{d\zeta_2}{ds} \approx -\vec{\omega}(\vec{\Delta}(\vec{z}))\cdot\hatvec{m}_0 \ .
\end{equation}

Defining $\vec{k}_0=\hatvec{l}_0-\mathrm{i}\hatvec{m}_0$, the first-order 1-turn spin-orbit coupling at some position $s_0$ is removed when

\begin{equation} \label{eq:s_o_integral}
\int_{s_{0}}^{s_{0}+C} \vec{\omega}(\vec{\Delta}(\vec{z}))\cdot\vec{k}_0 \, ds = 0 \ ,
\end{equation}
where $C$ is the total length of the closed orbit. Which $\vec{z}$ should we evaluate Eq.~\eqref{eq:s_o_integral} for? We can express any $\vec{z}$ as a linear combination of the symplectic orbital eigenvectors $\vec{v}_j$ in each plane $j=\pm I, \pm II, \pm III$. In a weakly-coupled ring, satisfying Eq.~\eqref{eq:s_o_integral} with $\vec{z}=\vec{v}_{\pm I}$ would be referred to as a ``horizontal strong synchro-beta spin match''. 


In real rings, random closed orbit distortions will, generally, reduce the polarization further (this is not always true; in fact, observing substantially better polarization with certain error seeds in the EIC-ESR caused the initial inspiration eventually leading to BAGELS). Even a slight tilt of $\hatvec{n}_0$ from the vertical in the arcs caused by closed orbit distortions may cause significant radiative depolarization, for example. In HERA, \textit{harmonic closed orbit spin matching} (HCOSM) was employed to correct this additional depolarization \cite{Barber:1985nd,Barber:1993ui}. HCOSM focuses on correcting the tilt of $\hatvec{n}_0$ from its design direction caused by the random errors, by using vertical orbit bumps that optimally cancel the integer Fourier harmonics of the tilt precession nearest to $\nu_0$. To quantify the tilt, a 1-turn periodic orthonormal coordinate system $(\hatvec{n}_0,\hatvec{m},\hatvec{l})$ is constructed in the \textit{ideal} ring where 

\begin{equation}
    \hatvec{m}(s)+\mathrm{i}\hatvec{l}(s) = e^{-\mathrm{i}\psi(s)}\left(\hatvec{l}_0(s)+\mathrm{i}\hatvec{m}_0(s)\right) \ ,
\end{equation}

and $\psi(s)$ is the so-called spin phasing function which must be chosen so that $\psi(s_0+C)-\psi(s_0)=2\pi\nu_0$ for 1-turn periodicity. For HERA, the spin phasing function was chosen to linearly evolve with $s$ so that 

\begin{equation}\label{eq:hera-spin-phase}
    \psi(s) = \psi(s_0)+2\pi\nu_0\frac{s-s_0}{C} \ .
\end{equation}

Assuming small closed orbit perturbations, various vertical orbit bumps in the ring are turned on and the original unperturbed $\hatvec{l}$ and $\hatvec{m}$ are attached to the new perturbed closed orbits. A Fourier transform of the projections of the new $\hatvec{n}_0$ on $\hatvec{l}$ and $\hatvec{m}$ for each can bump then be taken, and each bump's impact on the integer harmonics nearest to $\nu_0$ can be calculated. Finally, groups of the bumps are formed \textit{ab initio} to construct special bumps which ideally only impact a single integer harmonic of the $\hatvec{n}_0$ tilt along $\hatvec{l}$ or $\hatvec{m}$. These bumps are then varied to correct the depolarization caused by random closed orbit distortions.

While HCOSM did prove effective for HERA, it suffers from various disadvantages. Firstly, HCOSM of course is only intended to be applied for correcting the depolarization in a ring with random closed orbit distortions, not for increasing polarization in the ideal ring. Secondly, HCOSM indirectly attacks the real problem, which is $\vec{d}$: it assumes that tilts to $\hatvec{n}_0$ are always bad, and that bumps which maximally impact the integer spin harmonics nearest to $\nu_0$ will optimally correct the tilt and optimally impact $\vec{d}$. Thirdly, HCOSM as applied in HERA provided no systematic way of dealing with the spurious vertical dispersion created by the bumps, and coupling if there are sextupoles between the corrector coils in a bump. Fourthly, HCOSM does not ensure minimal orbit excursions for a maximum impact on the Fourier harmonics. As we show in Sec.~\ref{sec:bagels}, BAGELS suffers no such setbacks.

\section{Computational Methodology}\label{sec:methods}
Computation of Eqs.~\eqref{eq:Pdk},~\eqref{eq:taust}, and \eqref{eq:taudep} directly can be costly, and some useful approximations may be made to speed up the calculation and obtain a reasonable estimate. Firstly, instead of averaging over the ensemble, we can approximate by evaluating the integral only on the closed orbit $\vec{z}\co$, i.e. $\hatvec{n}\approx\hatvec{n}_0$, $\vec{d}\approx\vec{d}(\vec{z}\co)$, $\hatvec{\beta}\approx\hatvec{\beta}(\vec{z}\co)$, $\hatvec{v}\approx \hatvec{v}(\vec{z}\co)$, and  $\dot{\hatvec{v}}\approx \dot{\hatvec{v}}(\vec{z}\co)$. Secondly, using Eqs.~\eqref{eq:gamma} and \eqref{eq:beta} it is easy to show that in the ultra relativistic case $\vec{d}(\vec{z}\co)= \left.\gamma_0\frac{\partial\hatvec{n}}{\partial\gamma}\right\rvert_{\vec{z}\co}=\left.\frac{\partial\hatvec{n}}{\partial\delta}\right\rvert_{\vec{z}\co}$. This term can be easily extracted as a coefficient in the first-order truncated power series expansion of $\hatvec{n}$ around the closed orbit.  

These approximations give a generally excellent estimate for the Sokolov-Ternov terms (those NOT $\propto \vec{d}$). However, $\vec{d}$ can vary significantly even with small variations from the closed orbit. Nonlinearities in the orbital motion which excite a larger beam size may exacerbate the problem. Therefore, to accurately estimate the effects of radiative depolarization, nonlinear Monte Carlo spin tracking including radiation damping and radiation fluctuations is necessary. In such tracking spin-flip effects should be excluded and the initial beam distribution should be in radiative equilibrium. Then, $\tau_{dep}$ can be obtained directly from the slope of $P$ vs. $t$, and $P_{dk}$ can be approximated as

\begin{equation}
    P_{dk} \approx P_{st}\frac{\tau_{st}^{-1}}{\tau_{st}^{-1}+\tau_{dep}^{-1}} \ , 
\end{equation}

where $P_{st}$ is just $P_{dk}$ with $\vec{d}=\vec{0}$.

In this paper, we will refer to the approximation $\vec{d}\approx \vec{d}(\vec{z}\co)$ as the \textit{analytical} calculation because with this approximation the DK integral can be calculated quickly without tracking. When performing BAGELS, this approximation for $\vec{d}$ is used. Then for verification, nonlinear Monte Carlo tracking is performed. Here we track through 3$^\textrm{rd}$ order damped maps expanded around the closed orbit generated by the Polymorphic Tracking Code (PTC) \cite{Forest:573082}, between the centers of every bend in the lattice. At the bend centers, stochastic kicks simulating the effects of photon emission are applied to each particle. While the damped maps are not symplectic due to the map truncation, we expect the symplectic error to be minimal because there are separate maps used between every bend center in the ring. All tracking points include a bunch of 1000 particles, initialized to have the linear equilibrium emittances in each plane. After tracking for approximately four damping times so the beam is in nonlinear radiative equilibrium (2000 turns for the 18 GeV EIC-ESR), the nonlinear $\tau_{dep}$ is calculated from the following approximately 10 damping times (the next 5000 turns).

The Bmad accelerator software toolkit is used both for the analytical polarization calculations and the nonlinear tracking (as an interface to PTC) in this paper \cite{SAGAN2006356}. Routines to perform BAGELS were implemented in an open source Julia interface to Bmad's general purpose Tao program \cite{taojl}.

\section{BAGELS}\label{sec:bagels}
\subsection{Initial Approach}
As evident by the spin-orbit integral in Eq.~\eqref{eq:s_o_integral}, to achieve a spin match one must set $\vec{\omega}$ and/or $\hatvec{k}_0$ so the spin-orbit coupling cancels. Per Eq.~\eqref{eq:omega_b2}, $\vec{k}_0$ corresponds to $\hatvec{n}_0$ and the spin tune evolution around the ring (zeroth order spin precession on the closed orbit), while $\vec{\omega}$ corresponds to the first-order magnetic fields around the closed orbit (the optics) given the $\hatvec{k}_0$. Changes to $\hatvec{k}_0$ are usually viewed as challenging, as they are connected with the geometry of the ring (spin precession in the bends) and solenoid strengths. In the ESR, analytical conditions were derived using Eq.~\eqref{eq:s_o_integral} to achieve both a horizontal and longitudinal spin match outside the spin rotator \cite{vadimderivation,signorelli}. The horizontal spin match is achieved by choosing the optics in each solenoid module (setting $\vec\omega$), while the longitudinal spin match is achieved by both having a horizontal spin match and choosing special solenoid strengths (setting $\vec{k}_0$). However, the solenoid strength necessary for such is 11 Tesla, which is obviously infeasible. The lack of a longitudinal spin match excites a significant $\vec{d}$ around the ring.


Without any other systematic method to change $\hatvec{k}_0$, the only way to achieve some longitudinal spin match is to ``steal'' from part of the horizontal spin match by changing $\vec\omega$. Thus, we seek a method which minimizes the leftover $\vec{d}$ by intentionally tilting $\hatvec{n}_0$ near and through the rotator using vertical orbit bumps. The corresponding changes to $\vec{k}_0$ should accumulate spin-orbit coupling which perfectly cancels that leftover by the rotator. The orbit bumps therefore should be placed in the periodic FODO arcs directly around the spin rotator, and basically have a net effect of ``walking down'' the spin orbit coupling function to zero on either side of the rotator. Furthermore, the bumps must have essentially no impact on the optics around the ring. 

\begin{figure}[!t]
   \centering
   \includegraphics*[width=\columnwidth]{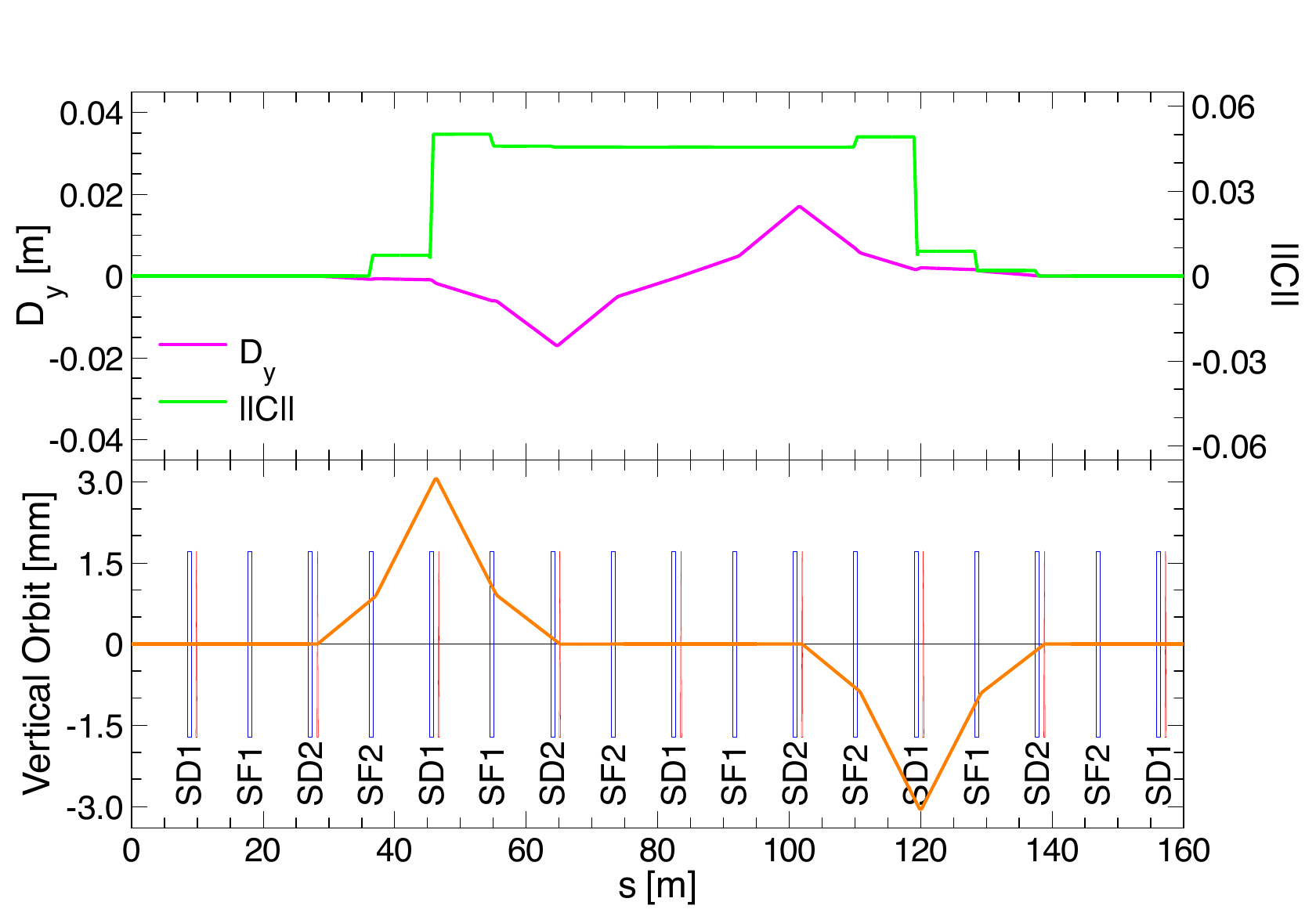}
   \caption{Opposite $\pi$ pair placed in a periodic FODO beamline with 90$\degree$ phase advance per cell and two sextupole families per plane (SF for the horizontal, SD for the vertical). Sextupoles are specified as blue rectangles, vertical coils as red triangles, and the quadrupoles and bends are not shown. The bump's corresponding vertical dispersion $D_y$ and normalized coupling matrix norm $||\mat{\bar C}||$ are plotted above \cite{PhysRevSTAB.2.074001}. The opposite $\pi$ pair creates only localized coupling and localized vertical dispersion, but a delocalized tilt to $\hatvec{n}_0$ due to the bending magnets between the coils, making it ideal for spin matching.}
   \label{fig:oppositepipair}
\end{figure}

An arbitrary choice of vertical orbit bump in the periodic arc section will create both delocalized vertical dispersion and delocalized transverse coupling (caused by the skew quadrupole feed-down term from a nonzero vertical closed orbit through the sextupoles). Both of these effects must, at least to first-order in the phase space coordinates, cancel by design of the bump. First, let us consider the coupling caused by the nonzero vertical orbits through the sextupoles. Because the chromatic beta-beat oscillates with $2\times$ the betatron oscillation frequency, chromatic sextupoles that are separated by 90$\degree$ in betatron phase will work against each other. Therefore, to coherently cancel the chromatic beta-beat, the sextupoles in a periodic arc are usually split into either two (for a 90$\degree$ betatron phase advance per cell) or three (for a 60$\degree$ betatron phase advance per cell) families per transverse plane, with each sextupole in a family separated by 180$\degree$ in betatron phase. We can take advantage of the chromatic sextupoles' regular placement to construct a bump which automatically cancels its own generated coupling. Consider the case of a 90$\degree$ phase advance per cell, as in the arcs of the 18 GeV EIC-ESR. If we turn on a $\pi$ bump, the orbit will through all four different strength sextupoles (two families per plane) once with a positive vertical offset. Therefore, in order to cancel this generated coupling, we need to use another $\pi$ bump of opposite strength such that the orbit goes through the four sextupoles again with the same magnitude but opposite sign. Therefore, we can place the second opposite $\pi$ bump anywhere $\pi n , \ n \in \mathbb{Z}$, in betatron phase away from the first bump in order to cancel the coupling. However, if we would also like to cancel the vertical dispersion wave generated by the first, then the first coil of the second $\pi$ bump must be $(2n+1)\pi$ away from the last coil of the first $\pi$ bump. 
This type of vertical orbit bump, which we henceforth we refer to as an ``opposite $\pi$ pair'' shown in Fig.~\ref{fig:oppositepipair}, has minimal impacts on the optics, creating no delocalized transverse coupling nor delocalized vertical dispersion. With bending magnets between the coils, a delocalized tilt to $\hatvec{n}_0$ is produced. This makes the opposite $\pi$ pair ideal for spin matching by intentionally changing $\hatvec{k}_0$. 


If the orbit excursions in the sextupoles are kept small, we can overlap all different opposite $\pi$ pairs in an arc and still have the orbit close and generate no delocalized vertical dispersion nor delocalized coupling. As such, we can construct special vertical orbit bumps, each composed of a linear combination of many opposite $\pi$ pairs, to minimize $\vec{d}$ in the bends where its contribution to radiative depolarization is most significant. Henceforth, we will refer to any choice of vertical orbit bump type which makes up a composite vertical orbit bump as a \textit{basis bump}. In this case, for spin matching, the basis bump choice is the opposite $\pi$ pair. 

If all bends in the ring have nearly the same length and strength, then considering only $\vec{d}$ at all bends is sufficient to capture the effects of radiative depolarization. More generally, per Eq.~\eqref{eq:taudep}, $\vec{d}\sqrt{L|\vec{g}|^3}$ at the ends of all bends could be considered, where $\vec{g}$ and $L$ are respectively the bend curvature and length. For the following formulation, we will assume all the bends have the same strengths and lengths for notational simplicity. Parametrizing the strength of each selected basis bump as $\theta_i$, the values of $\vec{d}$ at $r$ selected bends can be written as a function of $c$ selected basis bumps, to first-order in the strengths $\vec{\theta}=(\theta_1, ...,\theta_c)^T$, as

\begin{align}
\vec{f}_{\vec{d}} \approx(\vec{f}_{\vec{d}} )_{\vec{0}}+ \mat{R}_{\vec{d}}\vec{\theta}\label{eq:BAGELS} \ , 
\end{align}
\begin{align} \label{eq:BAGELS-matrix}\vec{f}_{\vec{d}} =\begin{pmatrix}
    \vec{d}_1 \\
    \vdots\\
     \vec{d}_r
\end{pmatrix}, \ 
   \mat{R}_{\vec{d}}=\begin{pmatrix}
        \frac{\partial\vec{d}_1}{\partial\theta_1} & \cdots & \frac{\partial\vec{d}_1}{\partial\theta_c} \\
        \vdots & \ddots & \vdots \\
        \frac{\partial\vec{d}_r}{\partial\theta_1} & \cdots & \frac{\partial\vec{d}_r}{\partial\theta_c}
    \end{pmatrix}_{\vec 0}  \ ,
\end{align}

where $\mat{R}_{\vec{d}}$ is the response matrix of $\vec{d}$ for each basis bump. Any arbitrary $\vec{\theta}$ defines a composite vertical closed orbit bump - a linear combination of the basis bumps - which can be scaled up and down using a knob in the control room of the real accelerator. We seek a $\vec{\theta}$ which sets the LHS of Eq.~\eqref{eq:BAGELS} closest to zero. Almost always, $3r > c$, and for this overdetermined system we can use any modern linear algebra software package to calculate a least-squares solution which minimizes $||\mat{R}_{\vec{d}}\vec{\theta}+(\vec{f}_{\vec{d}})_0||^2$. However, the problem with this approach is that the least-squares solution will have too large strengths for the basis bumps which have little impacts on $\vec{d}$ in the bends, exceeding the linear response regime and creating large orbit excursions. Naively, one could just iteratively exclude basis bumps that have the smallest columns in $\mat{R}_{\vec d}$. However, this approach is not general, as it is entirely possible, and in fact likely, that basis bumps which have large impacts alone may have little impact when paired together in a linear combination, and vice-versa. We thus require a systematic way to reduce our space to a minimal number $k$ of special $\vec{\theta}$'s which maximally scale when multiplied by $\mat{R}_{\vec{d}}$. Specifically, we seek

\begin{equation}\label{eq:RQ1}
    \argmax_{\vec{\theta}_1,...,\vec{\theta}_k}\frac{||\mat{R}_{\vec{d}}\vec{\theta}||^2}{||\vec{\theta}||^2} =  \argmax_{\vec{\theta}_1,...,\vec{\theta}_k}\frac{\vec{\theta}^T\mat{A}\vec{\theta}}{\vec{\theta}^T\vec{\theta}} \ , 
\end{equation}

where $\mat{A}=\mat{R}_{\vec{d}}^T\mat{R}_{\vec{d}}$ and $\vec{\theta}_1, ..., \vec{\theta}_k$ are all orthogonal to each other. These correspond to composite vertical orbit bumps which we can use to maximally, and orthogonally, impact the spin-orbit coupling function in the bends. Readers may recognize this well-known dimensionality reduction technique as principal component analysis (PCA). The quotient in Eq.~\eqref{eq:RQ1} is called the \textit{Rayleigh quotient} of $\mat{A}$ and $\vec{\theta}$, and we seek its maximizers. We follow the approach in \cite{ghojogh2023eigenvaluegeneralizedeigenvalueproblems}. It is obvious that we can express this problem equivalently as maximizing $\vec{\theta}^T\mat{A}\vec{\theta}$ subject to the constraint $\vec{\theta}^T\vec{\theta}=1$. Using the method of Lagrangian multipliers, the Lagrangian is

\begin{equation}\label{eq:lagr1}
    \Lagr = \vec{\theta}^T\mat{A}\vec{\theta} + \lambda\left(\vec{\theta}^T\vec{\theta}-1\right) \ .
\end{equation}

The stationary points of the Lagrangian are calculated as the solutions to

\begin{equation}
    \frac{\partial\Lagr}{\partial\vec{\theta}} = 2\mat{A}\vec{\theta} - 2\lambda\vec{\theta} = 0 \ \rightarrow\  \mat{A}\vec{\theta}=\lambda\vec{\theta} \ ,
\end{equation}
where we use the fact that, because the covariance matrix $\mat{A}$ is symmetric, $\partial_{\vec{\theta}} (\vec{\theta}^T\mat{A}\vec{\theta}) = 2\mat{A}\vec{\theta}$. Also, because $\vec{\theta}^T\mat{A}\vec{\theta}=||\mat{R}_{\vec{d}}\vec{\theta}||^2\geq 0$ for all $\vec{\theta}$ ($\mat{A}$ is \textit{positive semidefinite} by construction), all eigenvalues of $\mat{A}$ will be non-negative. Thus, the maximizers will be those eigenvectors $\vec{v}_i$ of $\mat{A}$ with the largest eigenvalues. These are the special linear combinations, or groups, of basis bumps which we can use to minimize the radiative depolarization in the real accelerator. In our reduced space of only $k$ knobs, we rewrite the least-squares problem as


\begin{align}
    -(\vec{f}_{\vec{d}})_0&=\mat{R}_{\vec{d}}(w_1\vec{v}_1 + ... + w_k\vec{v}_k) \nonumber\\
    \rightarrow \ -(\vec{f}_{\vec{d}})_0&=\mat{\tilde R}_{\vec{d}}\vec{w} \ ,\label{eq:BAGELS-LSQ}
\end{align}

where $\mat{\tilde R}_{\vec{d}}=\mat{R}_{\vec{d}}\begin{pmatrix} \vec{v}_1 & \hdots & \vec{v}_k \end{pmatrix}$ and $\vec{w}=(w_1,...,w_k)^T$.
In the ideal lattice, all we need to do now is calculate the least-squares solution to Eq.~\eqref{eq:BAGELS-LSQ} to get the strengths for each of the composite bumps which minimizes $\vec{d}$ in the bends. In the real ring where random closed orbit distortions will degrade the spin match, we could just vary the strengths of these $k$ bumps until the polarization is remaximized.

\subsection{Improving and Generalizing the Method}
By doing a principal component analysis over the space of selected opposite $\pi$ pairs in the ring, we thus far have a method to construct a minimal number of groups of basis bumps which have a \textit{maximal} impact on the spin-orbit coupling. However, at the same time we would also like the groups to have a \textit{minimal} impact on the orbit excursions, which the initial approach does not account for directly. Letting $\mat{R}_A$ specify the responses of the quantities we wish to maximally affect, and $\mat{R}_B$ specify the responses of those quantities we wish to minimally affect, we seek

\begin{equation}\label{eq:GRQ1}
    \argmax_{\vec{\theta}_1,...,\vec{\theta}_k}\frac{||\mat{R}_A\vec{\theta}||^2}{||\mat{R}_B\vec{\theta}||^2} =  
    \argmax_{\vec{\theta}_1,...,\vec{\theta}_k}\frac{\vec{\theta}^T\mat{A}\vec{\theta}}{\vec{\theta}^T\mat{B}\vec{\theta}} \ , 
\end{equation}

where $\mat{A}=\mat{R}_A^T\mat{R}_A$ and $\mat{B}=\mat{R}_B^T\mat{B}$. In this case, we would let $\mat{R}_A=\mat{R}_{\vec d}$ to maximally the spin-orbit coupling function at the bends, and $\mat{R}_B=\mat{R}_y$, where $\mat{R}_y$ is the response matrix of the vertical closed orbit at many positions around the ring, to minimally affect the closed orbit.

The quotient in Eq.~\eqref{eq:GRQ1} is called the \textit{generalized Rayleigh quotient} of $\mat{A}$, $\mat{B}$, and $\vec{\theta}$, and we seek its maximizers. We once again follow the approach in \cite{ghojogh2023eigenvaluegeneralizedeigenvalueproblems}. Assuming that $\mat{B}$ is symmetric and \textit{positive definite} such that $\vec{\theta}^T\mat{B}\vec{\theta} = ||\mat{R}_B\vec{\theta}||^2> 0$ for all $\vec{\theta}$, we can perform a Cholesky decomposition so that $\mat{B}=\mat{L}\mat{L}^T$. We then can rewrite the generalized Rayleigh quotient in Eq.~\eqref{eq:GRQ1} as

\begin{equation}\label{eq:cholesky}
    \frac{\vec{\theta}^T\mat{A}\vec{\theta}}{(\vec{\theta}^T\mat{L})(\mat{L}^T\vec{\theta})} = \frac{\vec{\tilde \theta}^T\mat{L}^{-1}\mat{A}\mat{L}^{-T}\vec{\tilde \theta}}{\vec{\tilde \theta}^T\vec{\tilde \theta}}\ , \ \  \vec{\tilde \theta}=\mat{L}^T\vec{\theta}\ .
\end{equation}

To calculate those $\vec{\tilde \theta}$ which maximize Eq.~\eqref{eq:cholesky}, the same Lagrangian as in Eq.~\eqref{eq:lagr1} is obtained except with $\vec{\theta}\rightarrow\vec{\tilde \theta}$. Expressing this in terms of $\vec{\theta}$, we have

\begin{equation}
    \Lagr=\vec{\theta}^T\mat{A}\vec{\theta}+\lambda\left(\vec{\theta}^T\mat{B}\vec{\theta}-1\right) \ .
\end{equation}

Therefore, the problem of Eq.~\eqref{eq:GRQ1} is equivalently expressed as finding the maximizers $\vec{\theta}^T\mat{A}\vec{\theta}$ subject to the constraint $\vec{\theta}^T\mat{B}\vec{\theta}=1$. The stationary points of the Lagrangian are calculated as the solutions to

\begin{equation}\label{eq:BAGELS-gen}
    \frac{\partial\Lagr}{\partial\vec{\theta}} = 2\mat{A}\vec{\theta} - 2\lambda\mat{B}\vec{\theta} = 0 \ \rightarrow\  \mat{A}\vec{\theta}=\lambda\mat{B}\vec{\theta} \ .
\end{equation}

In this case, the generalized eigenvectors of $\mat{A}$ and $\mat{B}$ with largest eigenvalues define groups of basis bumps which have a \textit{maximal} impact on the spin-orbit coupling, a \textit{minimal} impact on the orbit, and, by choice of the basis bumps, have no delocalized transverse coupling nor delocalized vertical dispersion. For spin matching purposes, these groups are truly the ``Best Adjustment Groups for ELectron Spin'' (BAGELS). We can use these BAGELS bumps to both optimally spin match the ideal ring via Eq.~\eqref{eq:BAGELS-LSQ}, and to optimally fix the spin match degradation caused by random closed orbit distortions in the real accelerator, by turning knobs which control these bumps in the control room. Being a more direct approach which also ensures minimal orbit excursions and no delocalized transverse coupling nor delocalized vertical dispersion, BAGELS effectively makes harmonic closed orbit spin matching obsolete. 

When calculating the generalized eigenvectors on a computer, floating point roundoff errors caused by the potentially vastly different scales of $\mat{R}_A$ and $\mat{R}_B$ (e.g. the closed orbit responses being orders of magnitude smaller than the spin-orbit coupling responses) may affect the numerical solution. Therefore in practice the two matrices should be normalized to minimize this error. We use the matrix 2-norm so that $\mat{\bar R}_A=\mat{R}_A/||\mat{R}_A||$ and likewise for $\mat{R}_B$. We then let $\mat{A}=\mat{\bar R}_A^T\mat{\bar R}_A$ and likewise for $\mat{B}$.

In this generalized approach, the positive definiteness of $\mat{B}$ is required in order to obtain a solution. This is apparent by the fact that if $\mat{B}$ is only positive semidefinite and not positive definite, then an attempt at inversion/Cholesky decomposition of $\mat{B}$ to solve for the generalized eigenvectors in Eq.~\eqref{eq:BAGELS-gen} will fail. In order to ensure $\vec{\theta}^T\mat{B}\vec{\theta} = ||\mat{\bar R}_B\vec{\theta}||^2>0$ for all $\vec{\theta}$
, all selected basis bumps must be orthogonal to each other - that is, no single selected basis bump should expressable as a linear combination of the other selected basis bumps. To construct an orthogonal basis for the opposite $\pi$ pair basis bump, we sequentially choose as our basis bumps all opposite $\pi$ pairs with a separation of only $\pi$ in betatron phase advance between each of the bumps in a pair. With this choice, those opposite $\pi$ pairs with separation $(2n+1)\pi,\ n\neq1$ can be constructed as a linear combination of our orthogonal basis bumps.

While we have thus far shown how BAGELS is used to calculate the best vertical orbit bumps for spin matching purposes, the method is not limited to spin matching. BAGELS can also be applied in applications where a \textit{minimal} impact on polarization is desired, amongst other things; suppose we instead would like to calculate bumps which maximally change some other quantity $x$ around the ring, but minimally impact both the spin-orbit coupling and the closed orbit excursions. To calculate such bumps, we can combine the responses of the spin-orbit coupling function and the orbit excursions into the response matrix $\mat{\bar R}_B$ (properly normalized), let $\mat{\bar R}_A=\mat{R}_x/||\mat{R}_x||$, and then choose those eigenvectors with the largest eigenvalues. This approach is used to calculate ``polarization-safe'' bumps to optimally correct global coupling in Sec.~\ref{sec:results:errors}, and ``polarization-safe'' vertical emittance creation bumps for beam size matching in Sec.~\ref{sec:results:emitb}. Finally, while it is not explored in this paper, the basis vectors of the ``groups'' do not need to be vertical orbit bumps, but could instead be single corrector coil strengths, or some other magnet strengths, for example. The only requirement is that a sufficiently linear relationship exist between the input and the output, and that the input forms an orthogonal basis. With such generality, BAGELS allows computation of the ``Best Adjustment Groups for ELectron Spin'' in any application.

In summary, to apply BAGELS:

\begin{enumerate}
    \item \textbf{Choose the basis bumps/vectors:} The choice of basis vectors should be guided by the goal. For spin matching, every orthogonal opposite $\pi$ pair, which create no delocalized vertical dispersion nor delocalized coupling, is ideal. In Sec.~\ref{sec:results:errors} and Sec.~\ref{sec:results:emitb} we define two different types of useful basis vectors for different applications.
    \item \textbf{Construct the response matrices:} All quantities in consideration must have a sufficiently linear relationship with the basis vectors. When combining the response matrices of multiple quantities into either $\mat{R}_A$ or $\mat{R}_B$, each submatrix may be normalized too to weigh the treatment of the different responses.
    \item \textbf{Calculate the generalized eigenvectors of the covariance matrices to obtain the Best Adjustment Groups for ELectron Spin:} The best groups will be those eigenvectors with the largest eigenvalues for maximal/minimal impact on polarization.
\end{enumerate}

\section{Results}\label{sec:results}
\subsection{Spin Match the Ideal Ring}
\subsubsection{1-IP 18 GeV EIC-ESR}\label{sec:results:1IP}

\begin{figure}[!t]
   \centering
   \includegraphics*[width=\columnwidth]{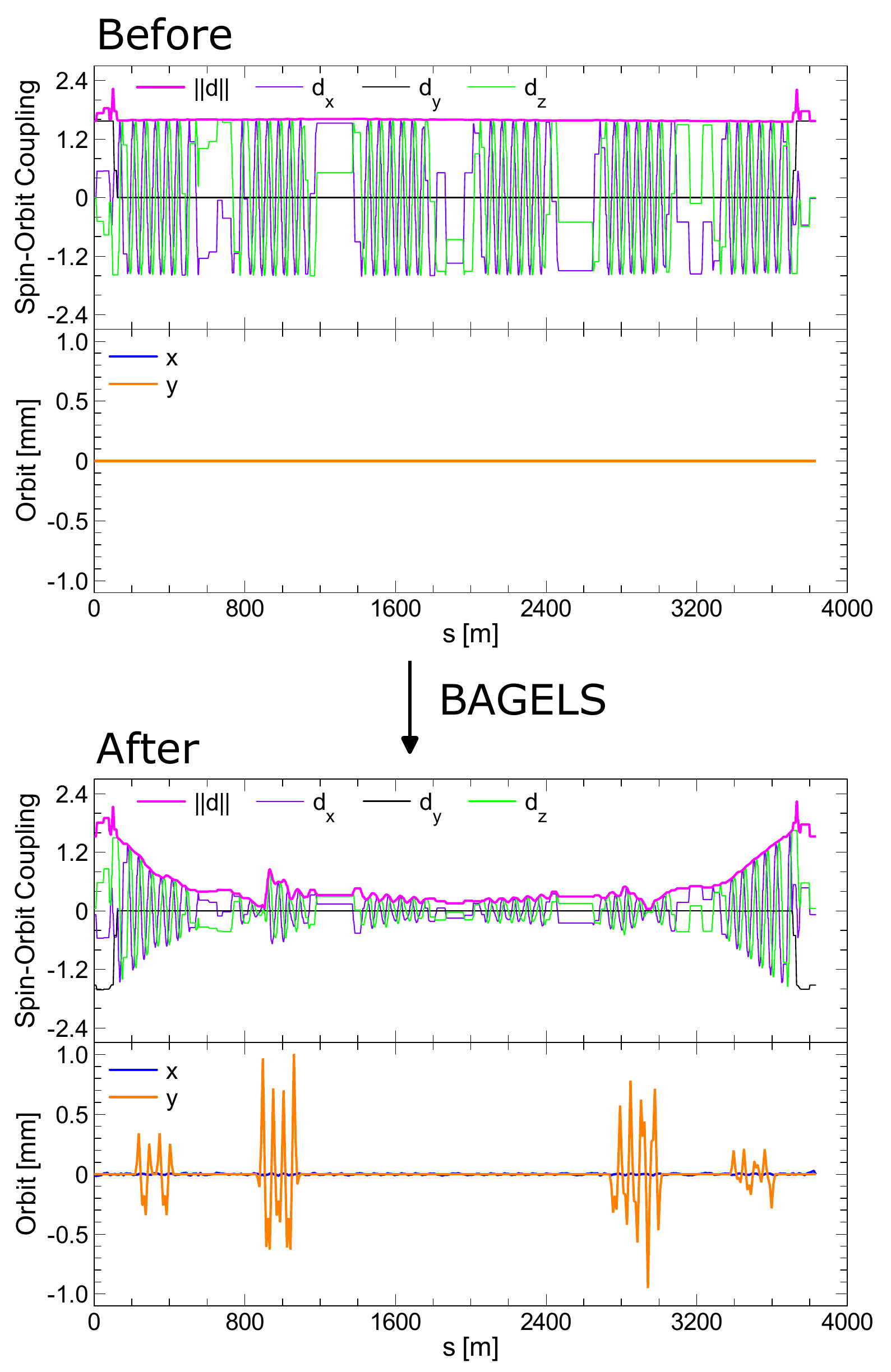}
   \caption{The spin-orbit coupling function $\vec{d}$ and closed orbit in the ideal 1-IP 18 GeV EIC-ESR before (top) and after (bottom) applying BAGELS, using only four BAGELS vertical orbit bumps. The interaction point at the center of the spin rotator shown in Fig.~\ref{fig:esr-rotator} is located at $s=0\ \textrm{m}$.}
   \label{fig:BAGELS}
\end{figure}

To apply BAGELS in the ideal 1-IP 18 GeV EIC-ESR, we use an orthogonal basis consisting of opposite $\pi$ pairs in the four arcs surrounding the spin rotator, two on either side. With radiation damping turned off, we calculate the responses of $\vec{d}\sqrt{L|\vec{g}|^3}$ at the ends of each bend (which we aim to maximize), as well as the vertical orbit $y$ at many positions around the ring (which we aim to minimize). Specifically, we sampled the ends of every lattice element (all drifts, quadrupoles, sextupoles, bends, solenoids) for $y$. Henceforth for notational brevity, $\mat{R}_{\vec d}$ denotes the response of $\vec{d}\sqrt{L|\vec{g}|^3}$ at all bends. We compute the top four maximizers of Eq.~\eqref{eq:GRQ1}, and use them as our knobs to achieve a spin match. Finally, in this reduced space of only four knobs, we solve Eq.~\eqref{eq:BAGELS-LSQ} for the least-squares solution to obtain the strengths of the four knobs in the ideal lattice. The before-and-after result is shown in Fig.~\ref{fig:BAGELS}, with the interaction point at $s=0\ \textrm{m}$.

The radiation damping will cause a ``sawtooth'' closed orbit, which must be corrected. Because we use the 1-IP ESR lattice for the random error studies in Sec.~\ref{sec:results:errors}, we use the proposed orbit correction scheme for the ring at the time of paper writing to correct the sawtooth closed orbit. This consists of dual plane BPMs and vertical corrector coils at each vertically-focusing quadrupole outside of the interaction region, and horizontal corrector coils at each horizontally-focusing quadrupole outside of the interaction region. In the interaction region, dual plane BPMs and dual plane correctors are used at each quadrupole. The RMS closed orbit was then minimized using the \textit{Tao} program in the \textit{Bmad} ecosystem. 
\begin{figure}[!t]
   \centering
   \includegraphics*[width=\columnwidth]{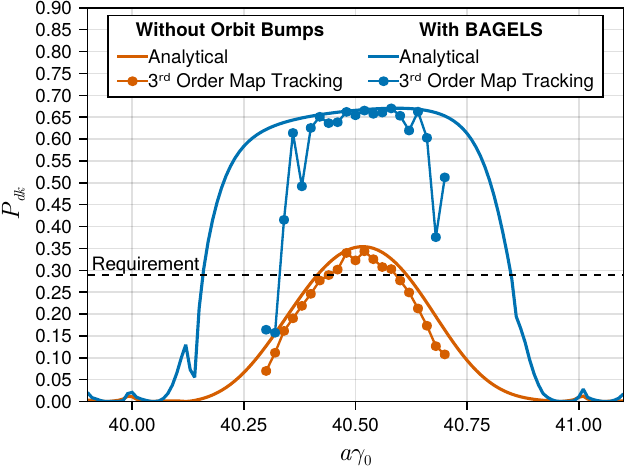}
   \caption{An energy scan of the asymptotic polarization in the ideal 1-IP 18 GeV EIC-ESR, both before and after applying BAGELS. Using only four BAGELS vertical orbit bumps, the asymptotic polarization is nearly doubled in nonlinear tracking.}
   \label{fig:1IP}
\end{figure}

Figure~\ref{fig:1IP} shows an energy scan of $P_{dk}$ of the 1-IP ESR before and after applying BAGELS. Note that polarized bunches will be injected into the ESR, and then replaced once sufficiently depolarized. The ``Requirement'' line shows the minimum allowable $P_{dk}$ in order to ensure a time-averaged polarization of $70\%$ is maintained. Using only four BAGELS bumps, which create a maximum orbit excursion of $\approx 1 \ \textrm{mm}$, we nearly double the asymptotic polarization.

We could have alternatively used only two of the arcs surrounding the IP instead of four, and similarly achieved a good solution. However, the orbit excursions must grow larger in order to solve Eq.~\eqref{eq:BAGELS-LSQ} for the least-squares solution, whereas with four arcs the orbit excursions are minimal.

\subsubsection{2-IP 18 GeV EIC-ESR}

\begin{figure}[!t]
   \centering
   \includegraphics*[width=\columnwidth]{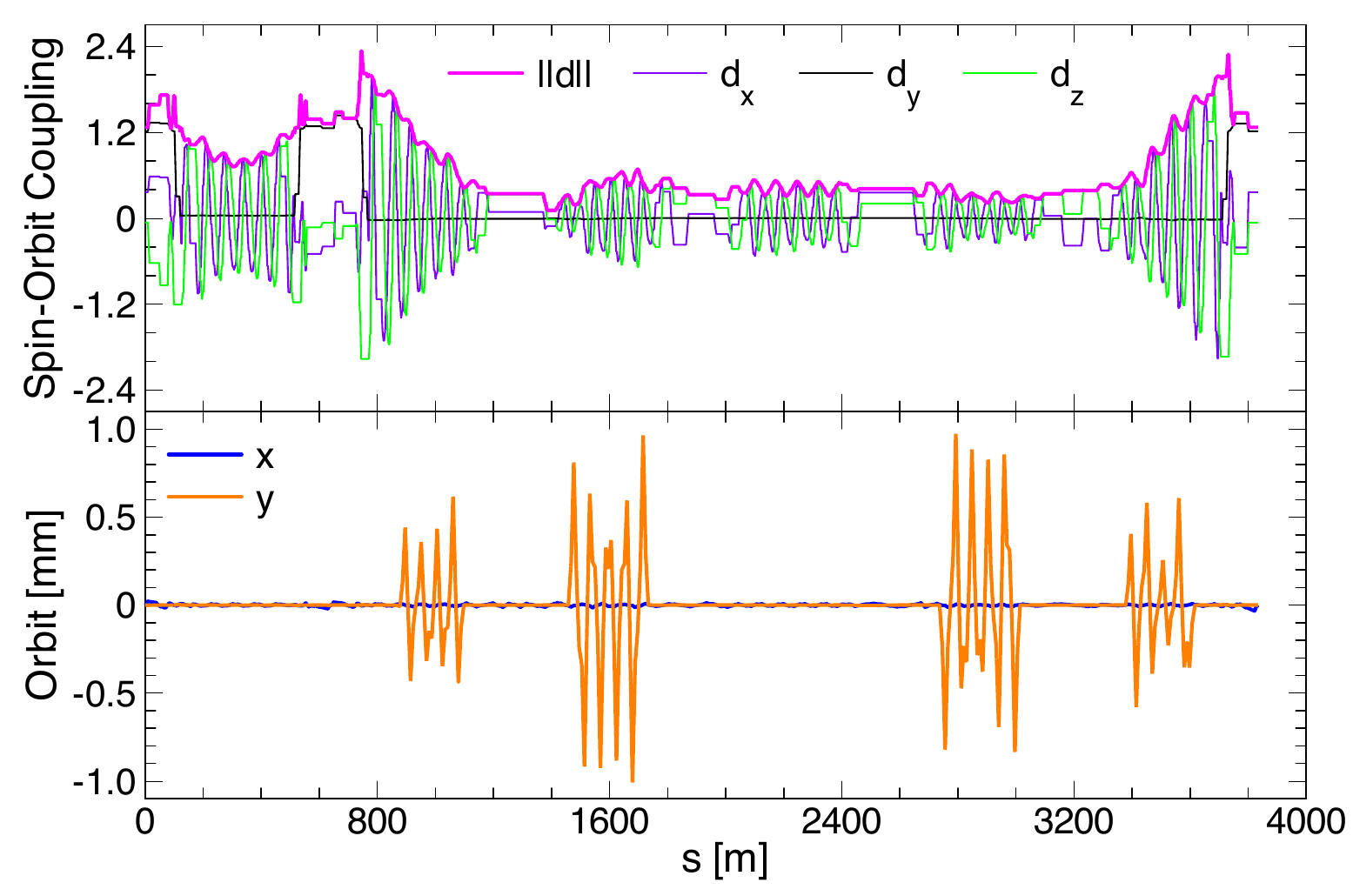}
   \caption{The spin-orbit coupling function $\vec{d}$ and closed orbit in the ideal 2-IP 18 GeV EIC-ESR before after applying BAGELS, using only four BAGELS vertical orbit bumps. The IPs at the center of the spin rotators (each shown in Fig.~\ref{fig:esr-rotator}) are located at $s=0\ \textrm{m}$ and $s=640.758\ \textrm{m}$.}
   \label{fig:2IP-BAGELed}
\end{figure}

In the 2-IP ESR, both interaction points including spin rotators are separated by one periodic arc section. To spin match this lattice, we use an orthogonal basis consisting of opposite $\pi$ pairs in the four arcs surrounding both spin rotators, i.e. two arcs upstream of the first spin rotator and two arcs downstream the second spin rotator. Once again applying BAGELS, with radiation damping turned off we calculate the responses of $\vec{d}\sqrt{L|\vec{g}|^3}$ at the ends of each bends and of the vertical orbit $y$ at the ends of each element. We compute the top four maximizers of Eq.~\eqref{eq:GRQ1}, and use them to solve for a least-squares solution to Eq.~\eqref{eq:BAGELS-LSQ}. Figure~\ref{fig:2IP-BAGELed} shows the spin-orbit coupling function and closed orbit after applying BAGELS to the ideal 2-IP 18 GeV ESR.

To correct the sawtooth closed orbit caused by radiation damping in the 2-IP lattice, we use the $\texttt{taper}$ command in the \textit{Tao} program; this command provides a quick-and-dirty solution by varying all magnet strengths in the ring proportionally to the local closed orbit momentum deviation. To ensure accuracy, we compared using \texttt{taper} vs. a proper orbit correction in the 1-IP lattice, and did not observe a significant difference in the tracking results.

\begin{figure}[!t]
   \centering
   \includegraphics*[width=\columnwidth]{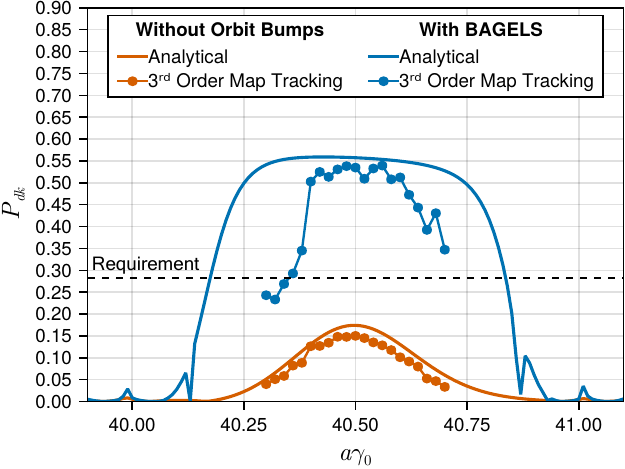}
   \caption{An energy scan of the asymptotic polarization in the ideal 2-IP 18 GeV EIC-ESR, both before and after applying BAGELS. Using only four BAGELS vertical orbit bumps, the asymptotic polarization is more than tripled in nonlinear tracking.}
   \label{fig:2IP}
\end{figure}

Figure~\ref{fig:2IP} shows an energy scan of $P_{dk}$ of the 2-IP ESR before and after applying BAGELS. Using only four BAGELS bumps, which create a maximum orbit excursion of $\approx 1 \ \textrm{mm}$, we more than triple the asymptotic polarization. Before BAGELS, the 2-IP 18 GeV ESR does not provide the sufficient time-averaged polarization at the interaction points \cite{Signorelli:2024xtq}. After application of BAGELS, the polarization requirements are exceeded.

\subsection{Random Errors and Global Coupling Correction}\label{sec:results:errors}

In the real ring, the BAGELS spin matching knobs can be used to optimally restore the spin match degradation caused by random closed orbit distortions. However before this is done, the global, delocalized transverse coupling caused by the random errors must be corrected. One way to do so is by using vertical orbits bumps through the sextupoles to create coupling which cancels out the coupling caused by the random errors. By applying BAGELS, we can compute a minimal number of such bumps which maximally, and orthogonally, impact the coupling, while minimally impacting the spin-orbit coupling function and creating minimal orbit excursions. 

We use the coupling formalism defined in \cite{PhysRevSTAB.2.074001}. The 1-turn transverse transport matrix is decomposed as

\begin{equation}
    \mat{M}_{4\times 4}=\mat{G}^{-1}\mat{\bar V}\mat{\bar U}\mat{\bar V}^{-1}\mat{G} \ ,
\end{equation}
\begin{align}
    \mat{G}=\begin{pmatrix}
        \mat{G}_a & \mat{0} \\
        \mat{0} & \mat{G}_b 
    \end{pmatrix}, \ \mat{\bar V}=\begin{pmatrix}
        \chi\mat{1} & \mat{\bar C} \\
        -\mat{\bar C}^+ & \chi\mat{1}
    \end{pmatrix},\    \mat{\bar U}=\begin{pmatrix}
        \mat{\bar U}_a & \mat{0} \\
        \mat{0} & \mat{\bar U}_b
    \end{pmatrix}, \nonumber
\end{align}

where $\mat{G}$ is a linear normalizing transformation (i.e. $\mat{G}_a$ and $\mat{G}_b$ are Courant-Snyder transformations up to a rotation of the ``a'' and ``b'' modes), $\mat{\bar V}$ is a symplectic matrix that defines the amount of coupling between the transverse planes (where the $+$ denotes the symplectic conjugate), and $\mat{\bar U}$ is an uncoupled rotation. In order to ensure the symplecticity of $\mat{\bar V}$, we demand $\chi^2+\det(\mat{\bar C})=1$.
If there is no transverse coupling, then $\chi=1$, $\mat{\bar C} = 0$, and the ``a'' mode corresponds to the horizontal and the ``b'' mode the vertical. The matrix $\mat{\bar C}$ is referred to as the \textit{normalized coupling matrix}, and provides a good measure of the coupling at a point in the ring.

\begin{figure}[t]
   \centering
   \includegraphics*[width=\columnwidth]{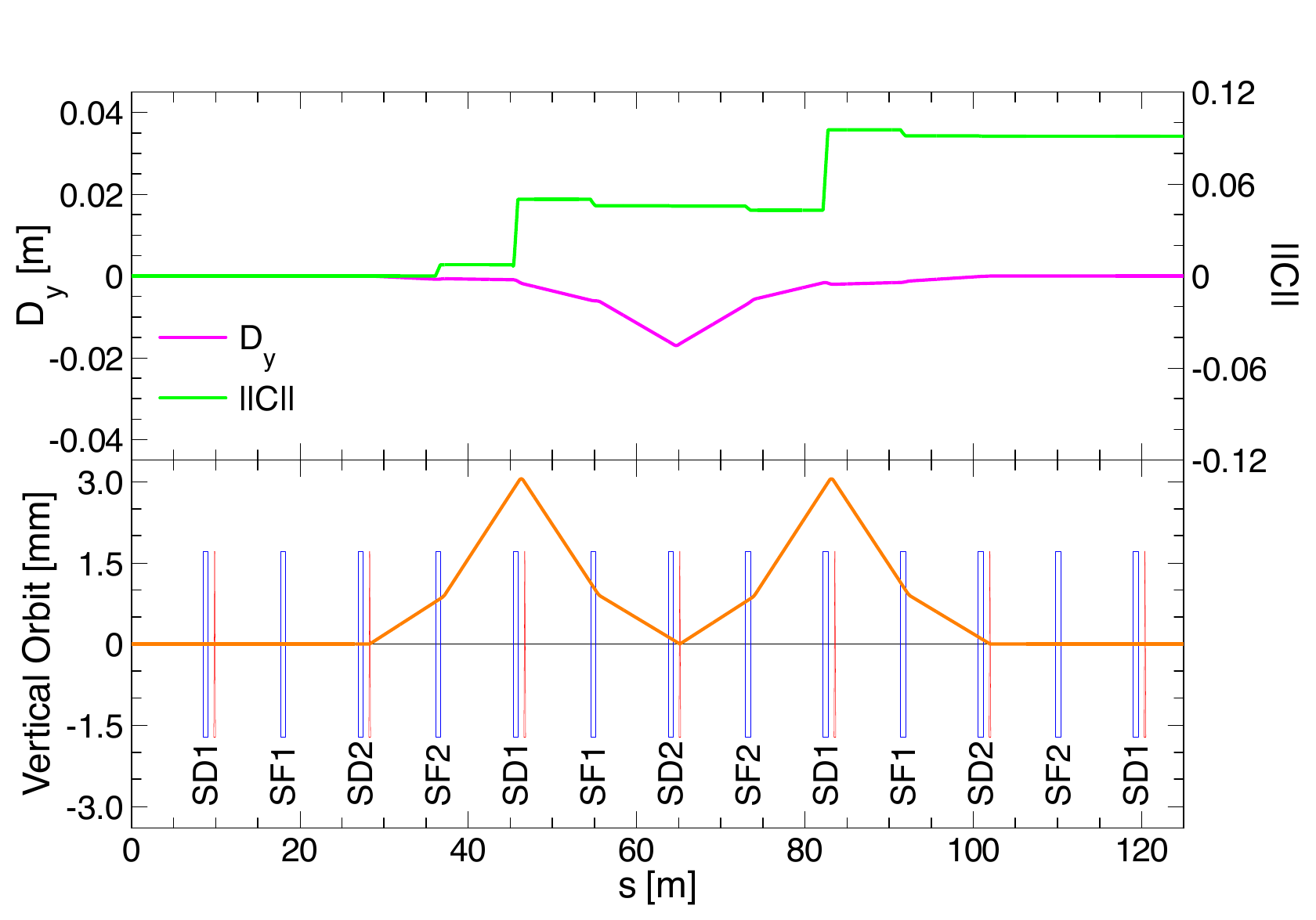}
   \caption{Equal $\pi$ pair placed in a periodic FODO beamline with 90$\degree$ phase advance per cell and two sextupole families per plane (SF for the horizontal, SD for the vertical). Sextupoles are specified as blue rectangles, vertical coils as red triangles, and the quadrupoles and bends are not shown. The bump's corresponding vertical dispersion $D_y$ and normalized coupling matrix norm $||\mat{\bar C}||$ are plotted above \cite{PhysRevSTAB.2.074001}. The equal $\pi$ pair creates delocalized transverse coupling but localized vertical dispersion.}
   \label{fig:equalpipair}
\end{figure}

We apply BAGELS seeking groups which maximally impact the normalized coupling matrix components $\vec{\bar{C}} = (\bar{C}_{11}, \bar{C}_{12}, \bar{C}_{21}, \bar{C}_{22})^T$ at each element in the ring, while minimally impacting both the spin-orbit coupling function at the bends and the orbit excursions at all elements. Following the recipe at the end of Sec.~\ref{sec:bagels}, we first must choose a basis bump. Ideally, we choose a bump which generates delocalized transverse coupling, but no delocalized vertical dispersion. Taking advantage of the periodic layout of the sextupole families in the arcs, two equivalent $\pi$ bumps directly next to each other, as shown in Fig.~\ref{fig:equalpipair}, is exactly such a basis bump. By sequentially choosing each equal $\pi$ pair in each arc, we obtain an orthogonal basis about which to construct our BAGELS global coupling correction knobs.

Next, we must construct the response matrices. Following the notation in Eq.~\eqref{eq:GRQ1}, we use

\begin{equation}
    \mat{R}_A = \mat{R}_{\vec{\bar{C}}},\ \mat{R}_B = \begin{pmatrix}
        \mat{R}_{\vec{d}}/||\mat{R}_{\vec{d}}|| \\
        \mat{R}_y/||\mat{R}_y||
    \end{pmatrix}
\end{equation}

to calculate bumps which maximally impact the coupling matrix components but minimally impact the both the orbit and the spin-orbit coupling function. Here, the submatrix normalizations serve an important purpose. The different responses may be multiplied by any arbitrary scalars to give a greater/lesser weight to those responses in the eigenvector calculation. To ensure equal treatment of all responses, we choose to normalize each submatrix by its 2-norm. This is now in addition to minimizing the the floating point roundoff errors caused by the potentially vastly different scales.

Finally, we compute the generalized eigenvectors and obtain the BAGELS bumps for global coupling correction via vertical orbits through the sextupoles. We chose only the top four BAGELS bumps for optimal, polarization-safe coupling correction. Knobs which control these bumps could be varied freely to correct the coupling, with minimal impacts on both the polarization and the closed orbit. BAGELS allows for the decoupling (no pun intended) of global coupling compensation and the polarization.

To test the efficacy of both the four BAGELS global coupling correction knobs and the four BAGELS spin matching knobs including random closed orbit distortions, we generated 10 1-IP ESR lattices with different error seeds using the RMS errors shown in Table \ref{tab:rms}.
\begin{table}[h]
\caption{\label{tab:rms}%
RMS Errors used in the Random Errors Study}
\begin{ruledtabular}
\begin{tabular}{@{} lccccc @{}}
 RMS Error & $x$ [mm] &$y$ [mm]  & Roll [mrad]  & $\Delta B/B$ [\%]  \\
  \colrule
  Dipoles & 0.2 & 0.2 & 0.5 & 0.1 \\
  Quadrupoles & 0.2 & 0.2 & 0.5 & 0.1 \\
  Sextupoles & 0.2 & 0.2 & 0.5 & 0.2 \\
  High-$\beta$ Dipoles & 0.2 & 0.2 & 0.5 & 0.05 \\
  Final Focus Quads & 0.1 & 0.1 & 0.5 & 0.05\\
\end{tabular}
\end{ruledtabular}
\end{table}

\begin{figure}[!t]
   \centering
   \includegraphics*[width=\columnwidth]{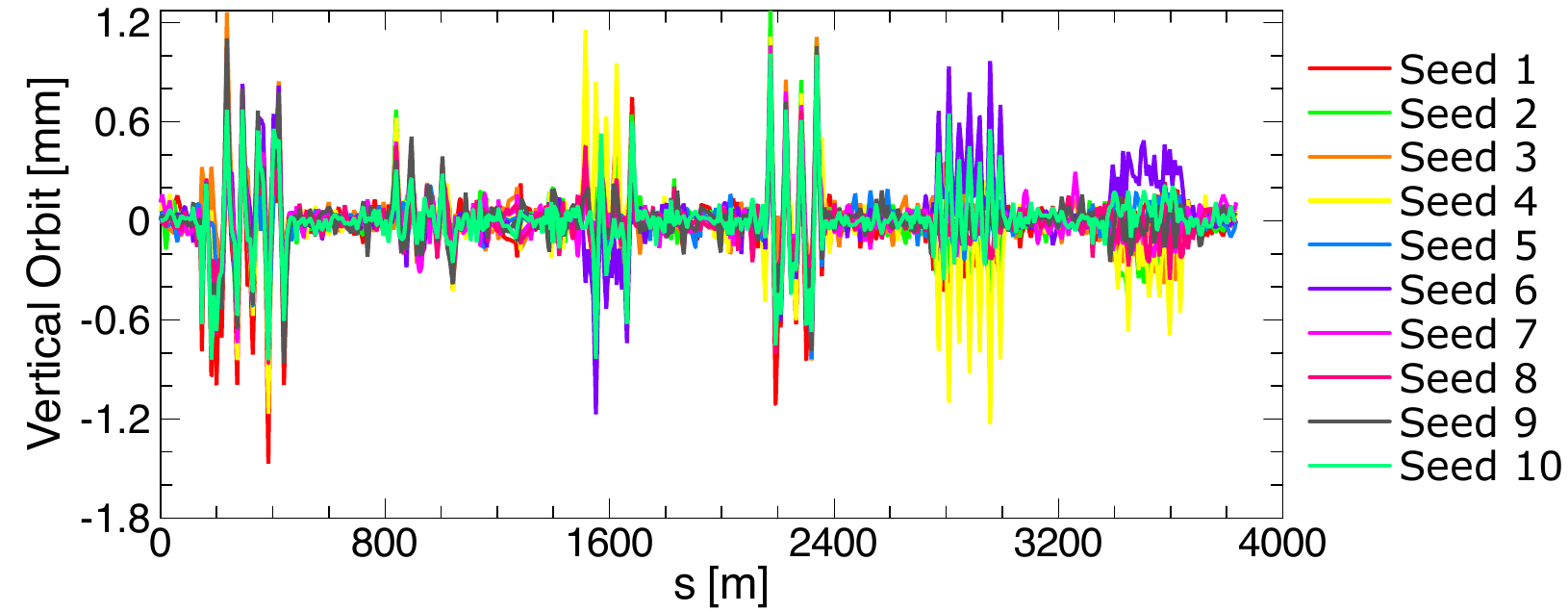}
   \caption{Vertical closed orbit in 10 error seeds of the 1-IP 18 GeV EIC-ESR after orbit correction, coupling correction using four BAGELS global coupling correction bumps, and spin match restoration using four BAGELS spin matching bumps. The IP is at $s=1277.948\ \textrm{m}$.}
   \label{fig:errors_y}
\end{figure}

\begin{figure}[!t]
   \centering
   \includegraphics*[width=\columnwidth]{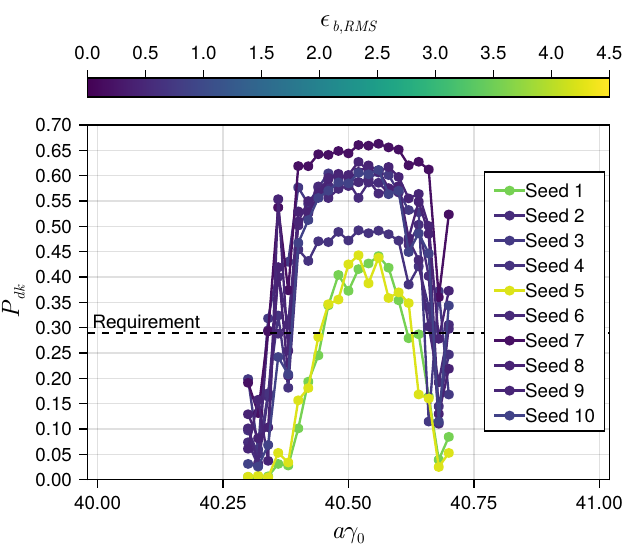}
   \caption{A 3$^{\textrm rd}$ order map tracking energy scan of the asymptotic polarizations in 10 different 1-IP 18 GeV EIC-ESR lattices including the RMS errors listed in Table~\ref{tab:rms}. Four BAGELS global coupling correction knobs were used to correct the coupling, followed by four BAGELS spin matching knobs to restore the spin match.}
   \label{fig:errors}
\end{figure}

 For generality, we use the eigenemittance calculation in \cite{WolskiPhysRevSTAB.9.024001} so that the ``horizontal-like'' emittance corresponds to $\epsilon_a$, and the ``vertical-like'' emittance to $\epsilon_b$. After first minimizing the RMS orbit for each seed using the orbit correction scheme outlined in Sec.~\ref{sec:results:1IP}, we varied the four BAGELS global coupling correction knobs until the RMS normalized coupling matrix components at each BPM was minimized. Then after resetting the tunes of the ring to their design values using the arc quadrupoles, the four BAGELS spin matching knobs were varied until the polarization was maximized. In all cases, the analytical (first order) $\epsilon_b$ were reduced to $<0.61 \ \textrm{nm}$. Figure~\ref{fig:errors_y} shows the vertical orbits for each seed after turning the BAGELS knobs, and Fig~\ref{fig:errors} shows an energy scan of the resulting asymptotic polarizations and mean $\epsilon_{b,RMS}$ obtained from 3$^\textrm{rd}$ order map tracking for each seed.

For all 10 error seeds, the polarization requirements for the 1-IP ESR are well exceeded. For two of the error seeds linear coupling correction does not sufficiently reduce the nonlinear vertical emittance. While emittance control is not the topic of this paper, it is remarkable that even in these cases BAGELS leads to sufficient $P_{dk}$ to achieve the polarization goals of the ESR. For 7 of the 10 seeds, asymptotic polarizations of $\approx60\%$ or more are achieved while having vertical equilibrium emittances in nonlinear tracking which agree very well with their linear calculations. For the two seeds with lower asymptotic polarization and a nonlinear $\epsilon_{b,RMS}$ increase, it is likely that a further optimization of the chromatic solution and/or harmonic sextupoles may gain control of the nonlinear $\epsilon_{b,RMS}$ and obtain a corresponding increase in $P_{dk}$. Overall, these results suggest that the four BAGELS spin matching bumps are highly efficacious and robust against random closed orbit distortions, and that the four BAGELS global coupling correction bumps can be used for excellent polarization-safe coupling control.

\subsection{Vertical Emittance Creation}\label{sec:results:emitb}
In order to achieve maximum luminosity in the EIC, the beam sizes of the electron and hadron beams must be matched. Because of radiation damping, the electron beam vertical emittance will be approximately zero, and if the vertical dispersion at the IP is to be kept to zero, then the vertical emittance must be intentionally increased in order to achieve a beam size match, specifically $\epsilon_y\approx2\ \textrm{nm}$. However, a nonzero vertical emittance is notoriously bad for electron polarization, and without a careful approach is all but guaranteed to significantly reduce the asymptotic polarization.

There are many different ways to generate vertical emittance. One way is to use a large vertical chicane as a local vertical dispersion bump, so that a vertical amplitude is excited when photons are emitted in the chicane. This was attempted in the ESR, however to achieve the necessary emittance the chicane requires too strong magnetic fields, geometrically is extremely challenging to fit in the ring, and degrades the spin match significantly \cite{Signorelli:2023fqc}. Alternatively, a localized region with a large amount of coupling where photons are emitted may be used to effectively ``couple'' part of the horizontal emittance into the vertical. However this has not been investigated in-depth, and will also certainly degrade the spin match significantly without a careful approach. 

Instead of a localized approach, vertical emittance could be created by using vertical orbit bumps that either (1) generate delocalized transverse coupling via the skew quadrupole feed-down term in the sextupoles, or (2) generate delocalized vertical dispersion, caused by the corrector coils themselves, so that photon emission causes a vertical amplitude excitation. We investigate both such approaches using BAGELS.

\subsubsection{Delocalized Transverse Coupling}\label{sec:results:emitb:c}

The BAGELS global coupling correction bumps we calculated in Sec.~\ref{sec:results:errors} have a maximum impact on the normalized coupling matrix components around the ring, a minimal impact on both the spin-orbit coupling function and the orbit, and by design create no delocalized vertical dispersion. This makes the BAGELS coupling correction bump with the largest eigenvalue perfect for creating vertical emittance via a delocalized transverse coupling wave. In the ideal 1-IP ESR, we simply turned this knob until the linear $\epsilon_b\approx\ 2\ \textrm{nm}$.

The working point of this ESR lattice is $(Q_x,Q_y,Q_s)=(0.08,0.14,0.05)$. Interestingly, we found in nonlinear tracking that delocalized coupling readily excited the $Q_y-Q_x-Q_s$ resonance, causing an uncontrolled vertical emittance increase. Because the correction of this effect is not the topic of this paper, we adjusted the working point slightly to $(0.08,0.15,0.045)$ by varying the arc quadrupoles and RF cavity voltages, to move away from this resonance.

\begin{figure}[!b]
   \centering
   \includegraphics*[width=\columnwidth]{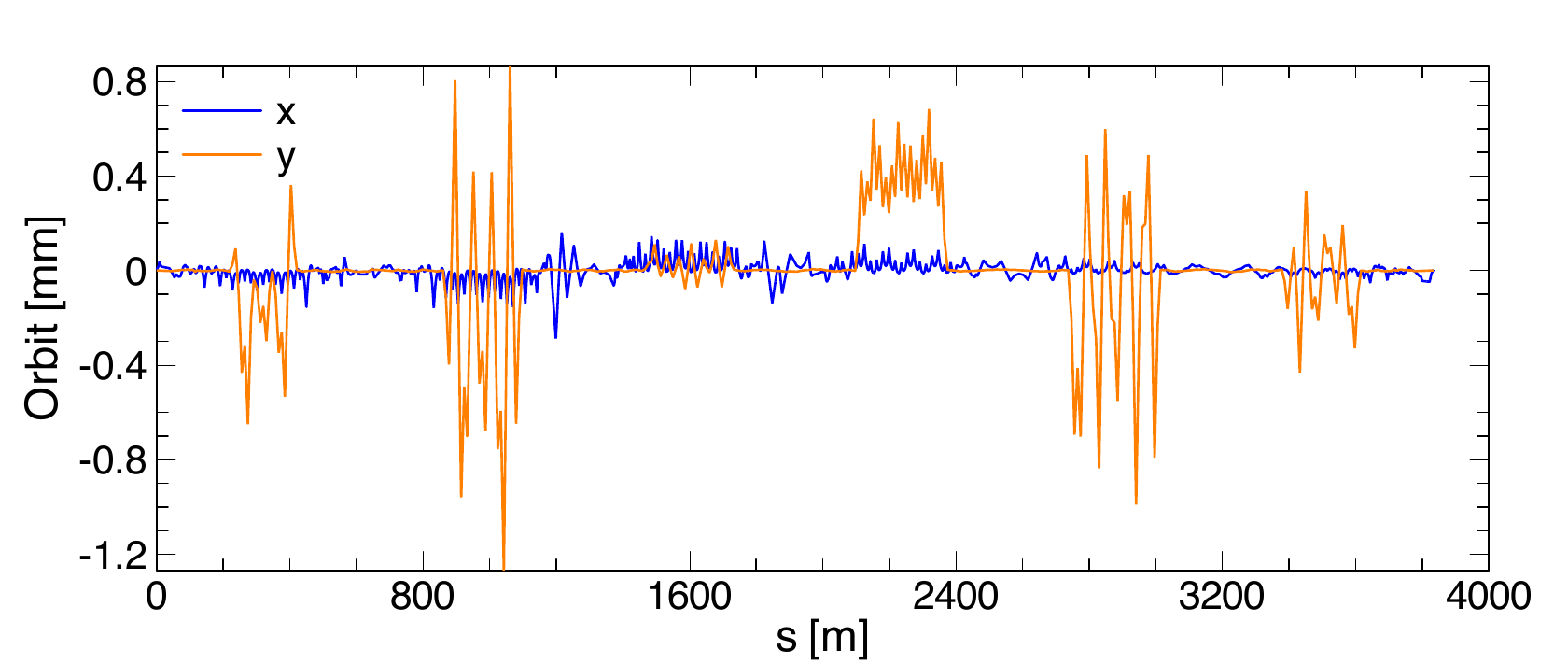}
   \caption{Closed orbit in a 1-IP 18 GeV EIC-ESR after using a single BAGELS single BAGELS coupling-creation bump to generate sufficient vertical emittance for beam size matching. The small horizontal orbit is due to radiation damping. The IP is located at $s=0\ \textrm{m}$.}
   \label{fig:c_y}
\end{figure}

We observed essentially zero impact on the analytical polarization after adjusting the BAGELS coupling-creating knob for linear $\epsilon_b\approx\ 2\ \textrm{nm}$; only miniscule adjustments to the four BAGELS spin matching bumps were needed to re-maximize the analytical polarization. The closed orbit after using the BAGELS bumps is shown in Fig.~\ref{fig:c_y}.

\begin{figure}[!t]
   \centering
   \includegraphics*[width=\columnwidth]{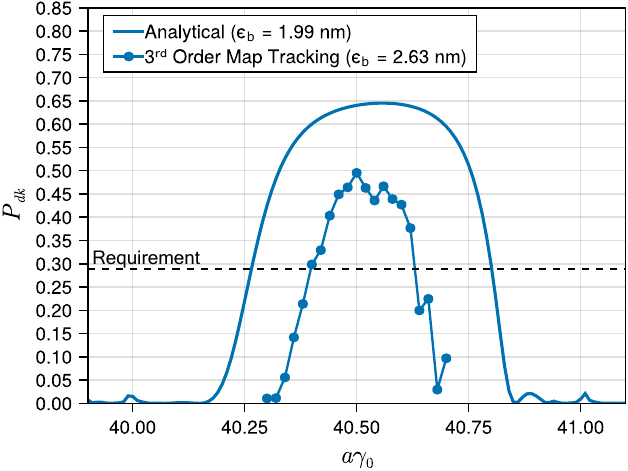}
   \caption{An energy scan of the asymptotic polarization in the ideal 1-IP 18 GeV EIC-ESR after using a single BAGELS coupling-creation bump to generate sufficient vertical emittance for beam size matching.}
   \label{fig:c}
\end{figure}

As shown in the energy scan in Fig.~\ref{fig:c}, using one BAGELS coupling-creation bump and four BAGELS spin matching bumps, polarization requirements are well exceeded while simultaneously achieving sufficient vertical emittance in nonlinear tracking, even with a small nonlinear increase in $\epsilon_b$. The disagreement of the analytical polarization calculation with nonlinear tracking is not surprising; the spin-orbit coupling function varies significantly with the vertical orbital amplitude, and the analytical calculation only evaluates the spin-orbit coupling function on the closed orbit.




\subsubsection{Delocalized Vertical Dispersion}
Using the coupling formalism as in Sec.~\ref{sec:results:errors}, we can define a- and b-mode periodic dispersions as

\begin{equation}
\begin{pmatrix}
    \eta_a \\ \eta_a' \\ \eta_b \\ \eta_b'
\end{pmatrix} = \mat{G}^{-1}\mat{\bar V}^{-1}\mat{G}\begin{pmatrix}
        \eta_x \\ \eta_x' \\ \eta_y \\ \eta_y'
\end{pmatrix} \ .
\end{equation}

The excitation of a ``b''-mode (vertical-like) amplitude caused by photon emission is directly characterized by

\begin{equation}
    \mathcal{H}_b= \bar{\eta}_{b1}^2 + \bar{\eta}_{b2}^2\ ,\ \ \bar{\eta}_{b1}=\frac{\eta_b}{\sqrt{\beta_b}}\ ,\ \bar{\eta}_{b2}=\frac{\alpha_b\eta_b+\beta_b\eta_b'}{\sqrt{\beta_b}}\ ,
\end{equation}

in places where photons are emitted \cite{Sands:1969lzn,SAGAN2006356}. The Twiss parameters are defined by the Courant-Snyder form of $\mat{G}_b$. Thus, to generate vertical emittance by radiation in areas where a vertical amplitude is excited, we seek a single BAGELS bump which maximally impacts $\mathcal{H}_b$ in the bends, but minimally impacts the spin-orbit coupling function in the bends and the orbit in all elements. The bump should also by construction generate no delocalized transverse coupling.

\begin{figure}[!h]
   \centering
   \includegraphics*[width=\columnwidth]{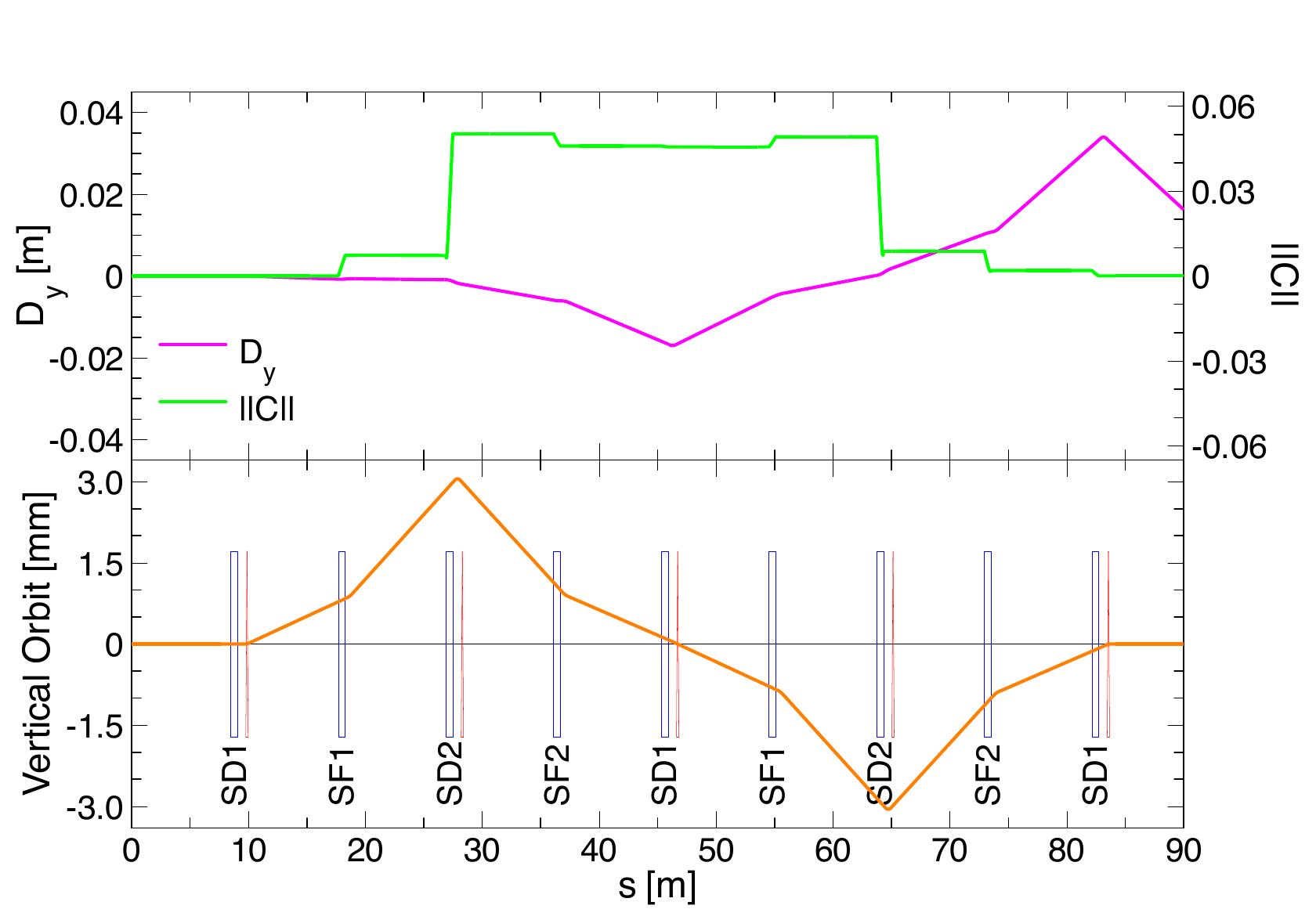}
   \caption{$2\pi$ bump placed in a periodic FODO beamline with 90$\degree$ phase advance per cell and two sextupole families per plane (SF for the horizontal, SD for the vertical). Sextupoles are specified as blue rectangles, vertical coils as red triangles, and the quadrupoles and bends are not shown. The bump's corresponding vertical dispersion $D_y$ and normalized coupling matrix norm $||\mat{\bar C}||$ are plotted above \cite{PhysRevSTAB.2.074001}. The $2\pi$ pair creates delocalized vertical dispersion but localized coupling.}
   \label{fig:2pibump}
\end{figure}

Once again, we follow the recipe at the end of Sec.~\ref{sec:bagels}. First, we need to choose a basis bump. To cancel the coupling created by a single $\pi$ bump in a periodic FODO section, we need to place an opposite $\pi$ bump position $\pi n$ away in betatron phase. For the ``opposite $\pi$ pair'' we used for spin matching, we positioned the second bump exactly $\pi$ away from the first, to cancel both the coupling and vertical dispersion. In this case, we want to coherently build vertical dispersion while cancelling the coupling, so instead we can place the second $\pi$ bump $2\pi n$ away from the first in betatron phase. To construct an orthogonal basis fully spanning the space, we position the second directly following the first so that $n=0$, and sequentially select all such combinations in each arc. We refer to this basis bump, shown in Fig.~\ref{fig:2pibump}, as a ``$2\pi$ bump'' for the obvious reason. The $2\pi$ bump, which generates delocalized vertical dispersion but no delocalized coupling, is the perfect basis bump to use for generating vertical emittance via delocalized vertical dispersion.

\begin{figure}[!t]
   \centering
   \includegraphics*[width=\columnwidth]{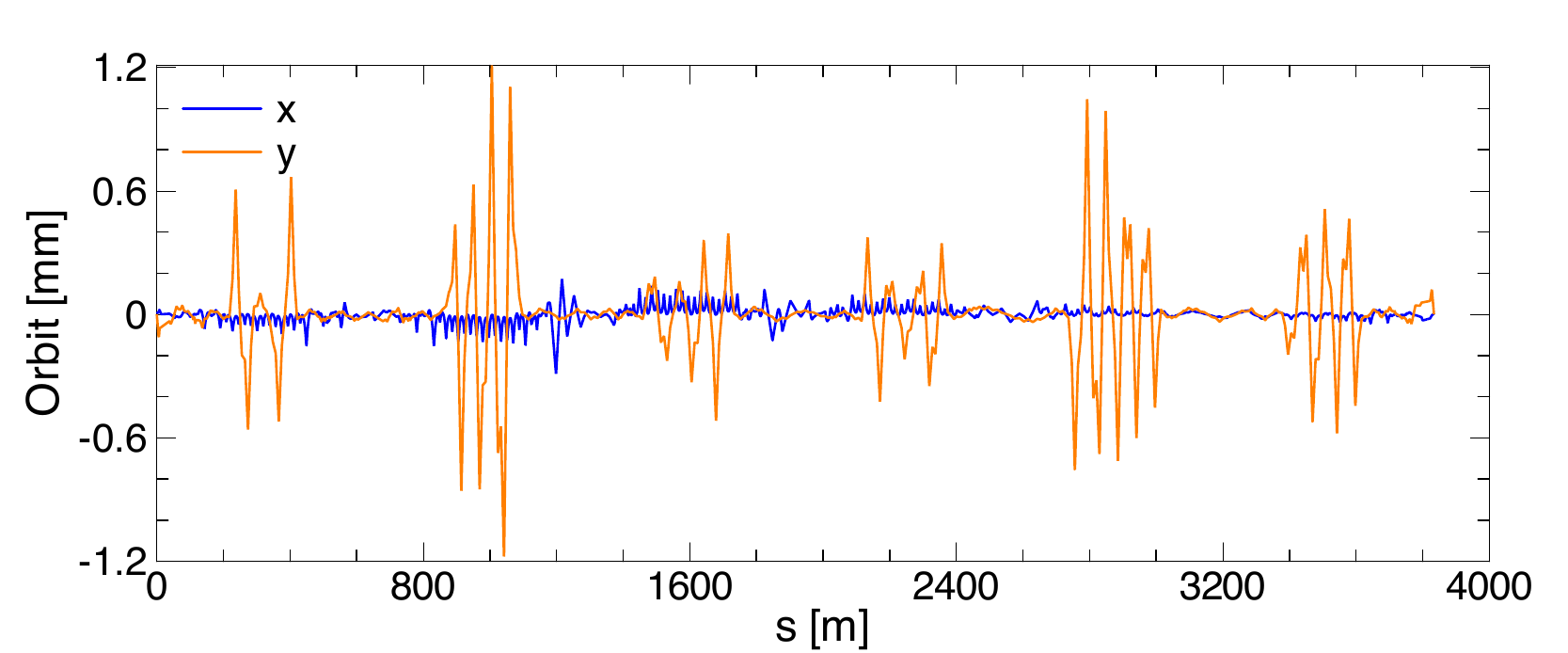}
   \caption{Closed orbit in a 1-IP 18 GeV EIC-ESR after using a single BAGELS vertical dispersion-creation bump to generate sufficient vertical emittance for beam size matching. The small horizontal orbit is due to radiation damping. The IP is located at $s=0\ \textrm{m}$.}
   \label{fig:etab_y}
\end{figure}

Now that we have chosen a basis bump, we must construct the response matrices. One of the requirements for BAGELS to work is a sufficiently linear dependence of the quantities with the basis vectors. In this case, $\mathcal{H}_b$ does not vary linearly with the $2\pi$ bump strength, however $\bar{\eta}_{b1}$ and $\bar{\eta}_{b2}$ do. Following the notation in Eq.~\eqref{eq:GRQ1}, for a maximal impact on $\bar{\eta}_{b1}$ and $\bar{\eta}_{b2}$ with a minimal impact on the spin-orbit coupling function $\vec{d}$ and the orbit $y$ we use

\begin{equation}
    \mat{R}_A=\mat{R}_{(\bar{\eta}_{b1},\bar{\eta}_{b2})}\ ,\ \mat{R}_B = \begin{pmatrix}
        \mat{R}_{\vec{d}}/||\mat{R}_{\vec{d}}|| \\
        \mat{R}_y/||\mat{R}_y||
    \end{pmatrix},
\end{equation}

to calculate bumps which maximally impact $\mathcal{H}_b$ and minimally impact the spin-orbit coupling and orbit. Finally, we use the one group with largest eigenvalue as our polarization-safe vertical dispersion creation bump, and vary that until $\epsilon_b\approx 2\ \textrm{nm}$ in the analytical calculation. Varying the BAGELS vertical dispersion creation bump caused essentially no changes to the polarization; only miniscule changes to the BAGELS spin matching knobs were made to re-maximize the analytical polarization. The closed orbit after using the BAGELS bumps is shown in Fig.~\ref{fig:etab_y}.

\begin{figure}[!t]
   \centering
   \includegraphics*[width=\columnwidth]{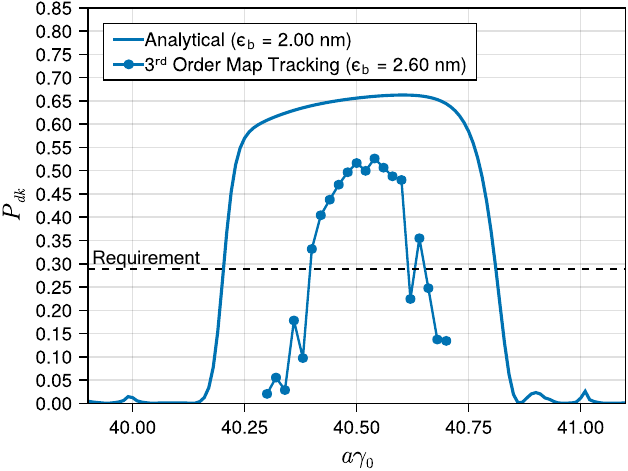}
   \caption{An energy scan of the asymptotic polarization in the ideal 1-IP 18 GeV EIC-ESR after using a single BAGELS vertical dispersion-creation bump to generate sufficient vertical emittance for beam size matching.}
   \label{fig:etab}
\end{figure}

Figure~\ref{fig:etab} shows an energy scan of the asymptotic polarization using the one BAGELS vertical dispersion creation bump for beam size matching. Excellent polarization, well exceeding the requirement, is achieved while simultaneously creating sufficient vertical emittance in nonlinear tracking. Just as with the delocalized coupling creation method in Sec.~\ref{sec:results:emitb:c}, only a small nonlinear increase in $\epsilon_b$ was observed, and with BAGELS the polarization requirements are still well-exceeded. Because the BAGELS vertical dispersion-creating bump does not require any change to the working point of the lattice to avoid nonlinear resonances, this method is recommended for polarization-safe beam size matching in the EIC-ESR.

\section{Conclusions}
Best Adjustment Groups for ELectron Spin (BAGELS) provides a rigorous framework to construct a minimal number of control parameter groups, or knobs, that have optimal impacts on the radiative depolarization, orbit, and optics. For spin matching purposes, we used BAGELS to compute special vertical orbit bumps, each a linear combination of basis bumps, that maximally affect the radiative depolarization around the ring while creating minimal orbit excursions and no delocalized vertical dispersion nor delocalized coupling. The BAGELS spin matching bumps work by intentionally tilting $\hatvec{n}_0$ to create spin-orbit coupling which can be used to cancel the spin-orbit coupling in the ring, e.g. spin rotators or random errors. For the design of the EIC-ESR, we used only four BAGELS spin matching knobs to significantly increase the asymptotic polarizations in nonlinear trackings of the 1- and 2-IP 18 GeV lattices, almost doubling $P_{dk}$ in the former and more than tripling $P_{dk}$ in the latter. For the EIC-ESR we also applied BAGELS in various applications where a minimal impact on polarization was desired: first, we calculated four BAGELS vertical orbit bumps which can be used to maximally correct the global transverse coupling caused by random errors, while minimally impacting the orbit and radiative depolarization. We then tested both these four BAGELS transverse coupling correction knobs and the four BAGELS spin matching knobs for 10 different error seeds in the 1-IP 18 GeV EIC-ESR, and showed excellent simultaneous polarization and transverse coupling control. Secondly, we calculated two different types of vertical emittance creation knobs - one which creates delocalized coupling, and one which creates delocalized vertical dispersion - that can be varied with truly minimal impacts to the radiative depolarization and orbit. Both knobs showed excellent simultaneous polarization and vertical emittance control in nonlinear tracking. BAGELS has solved the challenges of simultaneously maximizing polarization, correcting coupling, and creating the necessary vertical emittance ratio in the design of the EIC-ESR. These BAGELS knobs will be provided to the control room for operational optimizations. BAGELS will find similar utility in any present or future polarized lepton ring.

\begin{acknowledgments}
We thank Yunhai Cai and Yuri Nosochkov for useful input about coupling and dispersion canceling bumps, and Desmond P. Barber and David Sagan for many insightful conversations.
\end{acknowledgments}


\bibliography{apssamp}

\end{document}